\documentclass{aa}  

\usepackage[referable]{threeparttablex}
\usepackage{todonotes}
\usepackage{color}

\usepackage{longtable}
\usepackage{lscape}

\usepackage{graphicx}
\usepackage[normalem]{ulem}

\usepackage{txfonts}
\usepackage{soul}

\begin{document} 

\hyphenation{cata-logue}

   \title{Catalogue of central stars of planetary nebulae }

   \subtitle{Expanded edition}

%   \titlerunning{eeeeeeeeeee}
\authorrunning{Weidmann, et al.}

   \author{Weidmann, W. A. \inst{1}$^{,}$ \inst{2}, Mari, M. B. \inst{4}, Schmidt, E. O.\inst{3}, Gaspar, G. \inst{1}$^{,}$  \inst{2},
   Miller Bertolami, M. M. \inst{5}$^{,}$  \inst{6},
   Oio, G. A.\inst{3}, Gutiérrez-Soto, L. A.\inst{4} 
          Volpe M. G.\inst{3},  Gamen, R.\inst{5}$^{,}$\inst{6},  Mast, D.\inst{1}$^{,}$  \inst{2} }

   \institute{Observatorio Astron\'{o}mico de C\'{o}rdoba. Universidad Nacional de C\'{o}rdoba. Laprida 854, C\'{o}rdoba, Argentina\\
              \email{walter.weidmann@unc.edu.ar}
              \and
             Consejo de Investigaciones Científicas y Técnicas de la República Argentina, Argentina
         \and
             Instituto de Astronomía Teórica y Experimental. CONICET--UNC, Córdoba. Laprida 854, C\'{o}rdoba, Argentina. 
        \and
             Observatório do Valongo, Universidade Federal do Rio de Janeiro, Ladeira Pedro Antonio 43, 20080-090 Rio de Janeiro, Brazil
              \and
              Instituto de Astrofísica de La Plata, CONICET--UNLP, Buenos Aires, Argentina Paseo del Bosque s/n, FWA, B1900 La Plata, Buenos Aires, Argentina
              \and
              Facultad de Ciencias Astronómicas y Geofísicas, UNLP, Buenos Aires, Argentina Paseo del Bosque s/n, FWA, B1900 La Plata, Buenos Aires, Argentina
             }

%   \date{Received September 15, 1996; accepted March 16, 1997}

% \abstract{}{}{}{}{} 
% 5 {} token are mandatory

  \abstract
  % context heading (optional)
  % {} leave it empty if necessary  
  {
  Planetary nebulae represent a potential late stage of stellar evolution, 
  however the central stars (CSPNe) are relatively faint and therefore pertinent 
  information is merely available for $<$20\% of the Galactic sample. 
  Consequently, the literature was surveyed to construct a new catalogue of 
  620 CSPNe featuring important spectral classifications and information. 
  The catalogue supersedes the existing iteration by 25\%, and includes 
  physical parameters such as luminosity, surface gravity, temperature, 
  magnitude estimates, and references for published spectra. 
  The marked statistical improvement enabled the following pertinent 
  conclusions to be determined: the H-rich/H-poor ratio is 2:1, there 
  is a deficiency of CSPNe with types [WC 5-6], and nearly 80\% of 
  binary central stars belong to the H-rich group. 
  The last finding suggests that evolutionary scenarios leading to the 
  formation of binary central stars interfere with the conditions 
  required for the formation of H-poor CSPN.
 Approximately 50\% of the sample with derived values of 
$\log L$, $\log T_{\rm eff}$, and $\log g$,
exhibit masses and ages consistent with single stellar evolutionary models. 
The implication is that single stars are indeed able to form planetary nebulae. 
Moreover, it is shown that H-poor CSPNe are formed by higher mass progenitors. 
The catalogue is available through the Vizier database.
      }

%%%%%%%%%%%%%%%%%%%%%%%%%%%%%%%%%%%%%%%%%%%%%%%%%%%%%%%%%%%%%%%%%%%%%%%%%%%
%%%%%%%%%%%%%%%%%%%%%%%%%%%%%%%%%%%%%%%%%%%%%%%%%%%%%%%%%%%%%%%%%%%%%%%%%%%
%%%%%%%%%%%%%%%%%%%%%%%%%%%%%%%%%%%%%%%%%%%%%%%%%%%%%%%%%%%%%%%%%%%%%%%%%%%  


  \maketitle

%%%%%%%%%%%%%%%%%%%%%%%%%%%%%%%%%%%%%%%%%%%%%%%%%%%%%%%%%%%%%%%%%%%%%%%%%%%
%%%%%%%%%%%%%%%%%%%%%%%%%%%%%%%%%%%%%%%%%%%%%%%%%%%%%%%%%%%%%%%%%%%%%%%%%%%
%%%%%%%%%%%%%%%%%%%%%%%%%%%%%%%%%%%%%%%%%%%%%%%%%%%%%%%%%%%%%%%%%%%%%%%%%%%

\section{Introduction}

Catalogues are an important tool in astrophysics.
Numerous are the works in which the catalogues have helped researchers in the past.
For example, \cite{2007PASP..119.1349M} looked for a link between planetary nebulae (PNe) an open cluster, 
 which led in part to the discovery of PN PHR~1315$-$6555 in the cluster Andrews-Lindsay 1 and a 4\% distance solution \citep{2014A&A...567A...1M}.
\cite{2009A&A...496..813M} discovered 21 periodic  binaries at the center of planetary nebulae 
by cross-matching PNe and OGLE microlensing survey catalogues. They also established that the close binary fraction was 10-20\% in a independent sample of binary PNe and, in a follow up work \citep{2009A&A...505..249M}, they concluded that a fraction of at least 60\% of post-common-envelope PNe are bipolar nebulae. Thus changing the fact that NGC~2346 being the only post-common-envelope canonical bipolar.
More recently, \cite{2020ApJ...889...21S} matched the astrometry of PNe with DR2 Gaia parallaxes, thus providing the parallax to 430 objects. 
 Using these parallaxes they located the CS of the PNe and, together with effective temperatures and magnitudes from the literature, determined their masses.
The catalogs were the starting points for some important determinations.

Regarding the central star of planetary nebulae (CSPNe),
a complete catalogue but with limited information is
\cite{2011A&A...526A...6W}.
It has proven to be a useful tool for the astronomical community. But as during 
the last nine years there has been an increase in the number of CSPNe with 
new determined spectral types 
\citep[or improved through better quality/resolution spectra as][]{2018A&A...614A.135W},
an upgrade of that catalogue becomes necessary.
In addition, during these years many new developments have been published concerning the CSPNe.
There are spectral types that no longer apply, for example the wels \citep{2015A&A...579A..86W}.
Moreover, several [WN] stars could be confirmed, i.e. IC~4663 \citep{2012MNRAS.423..934M}, 
A~48, and probably at PB~8 \citep{2010A&A...515A..83T}.
Also NGC~5189, a binary system whose primary star is of the [WR] type \citep{2015MNRAS.448.1789M}. 
With regard to the multiplicity of the CSPNe,
it was possible to identify a triple system at the nucleus of the planetary nebula NGC~246
\citep{2014MNRAS.444.3459A}.
This situation, which is speculated to be common, 
might happen in LoTr~5 and SuWt~2.

 A catalogue would allow statistical analysis and provide observational 
 limits to evolutionary models, such as the progenitor's mass of PNe.
Although the range of masses of the progenitor star
which would result in a PN is frequently estimated to be between 1-8 M$_{\odot}$, 
the observational evidence in support of this point is  limited.
Nevertheless, the confirmation of a PN associated to an
open cluster, has a direct implication on the upper limit on the progenitor's mass.
\cite{2019MNRAS.484.3078F} and \cite{2019NatAs.tmp..357F}
have proved that the PNe  
PHR~1315$-$6555 and BMP~J1613$-$5406
are physically linked to an open cluster.
Consequently we have, for the first time, observational evidence 
 that the upper limit of the initial stellar mass for the formation of a 
 PN can be as high as M$_{*}$$\lesssim$5.0 M$_{\odot}$.
Another important step forward in the study of the formation of PNe 
came from the study of Hen~2$-$428, whose central star was found by 
\cite{2015Natur.519...63S} to be composed of a double-degenerate core 
with a combined mass above the Chandrasekhar limit. 
Although later studies have put such high mass into question, 
favouring a much lower  mass for the central stars 
\citep[e.g.][]{2016NewA...45....7G, 2018Galax...6...88R}. 
Specifically, the detailed analysis of \cite{Reindal} 
indicates that the system might be composed of a 0.66 M$_\odot$ 
post-AGB and a reheated 0.42 M$_\odot$ post-RGB star.
The study of this peculiar object will certainly shed light on 
the formation process of PNe through common envelope events.

 To establish an evolutionary sequence involving the spectral types of the CSPNe  
 was not a simple task, and certainly,  we  still have a lot to learn about this topic.  
 Proof of this is the case of Hen~3$-$1357.
 \cite{2017MNRAS.464L..51R} observed that the CSPN of this object is evolving  quickly, changing from  B-type ($T_{\rm eff}\sim21$ kK) object in 1971 to a $T_{\rm eff}\sim60$ kK peak in 2002. Since then it seems to be cooling and expanding back to the AGB at a rate of $770\pm580$ K yr$^{-1}$, similar to that of FG Sge \citep[$350$ K yr$^{-1}$][]{2006A&A...459..885J} and providing further evidence of the existence of late helium 
 flashes in the post-AGB evolution.
 
In summary, in the last years a series of extreme phenomena has 
been found in PNe that increase the complexity of these objects. 
Undoubtedly, the less understood stage of stellar evolution for low 
and intermediate mass stars is precisely the last one, i.e. from post-AGB to pre-WD.
This is the main motivation to update and expand the catalogue of CSPNe. 
Thus, in this work we present a new catalogue of spectral types of 
CSPNe which supersedes the \cite{2011A&A...526A...6W}.

This paper is structured as follows: 
In Section~\ref{st} we present a summary of the different spectral types of CSPNe. 
An evolutionary sequence is described in Section~\ref{revestE}. 
We detail the main body of the catalogue in Section~\ref{catal}. 
Results are presented in Section~\ref{reul}, 
while in Section~\ref{conclu} there is a summary of this work and we lay our conclusions.
Finally, in the Appendix we focus on peculiar CSPNe.

%%%%%%%%%%%%%%%%%%%%%%%%%%%%%%%%%%%%%%%%%%%%%%%%%%%%%%%%%%%%%%%%%%%%%%%%%
%%%%%%%%%%%%%%%%%%%%%%%%%%%%%%%%%%%%%%%%%%%%%%%%%%%%%%%%%%%%%%%%%%%%%%%%%
%%%%%%%%%%%%%%%%%%%%%%%%%%%%%%%%%%%%%%%%%%%%%%%%%%%%%%%%%%%%%%%%%%%%%%%%%

\section{Review of spectral classification}\label{st}

 It is well known that the CSPNe are divided into two large groups. 
 Those that present hydrogen absorption lines (H-rich) and those that do not (H-poor). 
 One of the pioneering work in compiling CSPNe spectral information was \cite{1975MSRSL...9..271A}. 
 He reported the following types: O-type, WC, OVI (currently called [WO]), Of + WR, sdO and continuum.

There are few spectral types that are specific of CSPNe i.e. O(C), 
[WC]-PG~1159, and Of-WR(H). For the [WR], O and Of-type, the classification 
criteria for massive WC stars (or massive O-type stars) is used. 
Implementing an appropriate criteria for CSPNe requires a large number of good quality spectra.

It is appropriate to point out that  
this is a qualitative classification system in which the 
unknown stellar spectrum is compared with a grid of standard stars.
In this sense, \cite{1999RvMA...12..255D} and
\cite{1998RvMA...11....3N}
present a collection of spectra of different spectral types of evolved objects
which represents a  useful tool.
In Table~\ref{lineSPT} we 
 mention some key lines of each spectral type.
 Nevertheless, these do not replace the proper  
   classification criteria of every type.
Below we present a guide, with bibliographic references, 
that describes the spectrum (in the optical range) of each type of object mentioned in our catalogue.

\begin{itemize}
\item $[WR]$ (subtype [WN], [WC] and [WO]):
\cite{1998MNRAS.296..367C},
\cite{2003A&A...403..659A} and
\cite{2012MNRAS.423..934M}.
The spectral type VL was implemented by
\cite{2009A&A...500.1089G}, which is in fact consistent with a late [WC].
\item O(H): Although these objects
follow the classification criteria for massive O-type
stars, in 
\cite{2018A&A...614A.135W}
some characteristics of the
O-type stars 
that are CSPNe
 are described.
In this type of stars it is possible to improve the
spectral classification with qualifiers
\citep{2011ApJS..193...24S,2014ApJS..211...10S,2016ApJS..224....4M}.
Based on the ions identified in the spectrum,
the CSPNe could be classified as O(H), O or H-rich
\citep{2018A&A...614A.135W}.
\item Of(H): This includes the 
O-type stars with emission lines.
Historically, there was a distinction between the O-type and Of-type stars.
Nevertheless, not all O-type star with emission lines are in fact an Of-type. 
To clarify this see
 \cite{1990A&A...229..152M,1969ApJ...157.1245S}.
 In  \cite{2011ApJS..193...24S} these objects are indicated with the
 qualifiers: f, f*, fc and f+.
\item O(C):
This type of star appears only in one member, i.e. K~3$-$67.
Previously, 
Lo~4, wray~17$-$1, and IsWe~1 were included in this spectral type
\citep{1991IAUS..145..375M}.
Nowadays, these last three objects were reclassified as PG~1159.
\item O(He):
\cite{2014A&A...566A.116R},
\cite{2008ASPC..391..135R},
\cite{2006ASPC..348..194R} and
\cite{1998A&A...338..651R}.
\item Of$-$WR(H):
\cite{1990A&A...229..152M} and \cite{1991IAUS..145..375M}.
\item WD (subtype DA, DAO and DO):
\cite{1999A&A...350..101N},
 \cite{1987ApJS...65..603M}, and
       \cite{1999ApJS..121....1M}.
\item hgO(H):
The classification hgO(H) was used exclusively for 
CSPNe \citep{1986ASSL..128..323M},
these objects are essentially evolved objects like
 sdO or WD stars.
 For example Sh~2$-$68 
 have three classifications i.e.
hgO(H), WD, and PG~1159
\citep{1995A&A...301..545N}.
\item sdO, sdB:
\cite{1997A&AS..125..501J} and
\cite{2019ApJ...882..171A}.
\item PG~1159 (subtype A, E and lg E):
\cite{1992LNP...401..273W},
\cite{1998A&A...334..618D} and
\cite{2005A&A...442..309H}.
\item $[$WC$]-$PG~1159:
Previously called Of-WR(C)
\citep{1991IAUS..145..375M}.
There are only two objects with this spectral type, 
A~30 and A~78.
\item hybrid:
\cite{1996A&A...309..820D} and
\cite{1999A&A...350..101N}.
\item symbiotic star (SySt):
This is not strictly a spectral type
(see Appendix~\ref{sec-syst}).
\item B[e]: 
\cite{2017ASPC..508....3O}
and 
\cite{1998A&A...340..117L}.
\item cont.:
\cite{2018A&A...614A.135W}.
This is not strictly a spectral type.
In general, objects with this designation are O-type stars.
\item B2-M9:
Objects later that a B1  does not have the required temperature 
to ionize the nebula.
In this sense, 
objects classified with a late type are, in fact, candidates for binary 
systems whose hot component is hitherto undiscovered. 
\item wels: Classification implemented by
\cite{1993A&AS..102..595T}.
This designation has been shown not to correspond to an independent spectral type. 
In general it contemplates  hot O(H)-type stars
\citep{2015A&A...579A..86W}.
\item EL (Emission Line): Few objects in which the presence of
emission lines of stellar origin have been reported. 
These objects do not always satisfy the wels criterion.
\item Blue: The MASH catalogue
identified CSPNe of some objects.
Photometric observations indicate that these objects are blue.
None of these objects have spectroscopic information.
\end{itemize}

\begin{table*}[ht]
\caption{Key optical lines to spectral classification of CSPNe. 
Upper panel represents narrow features, and lower, wide features.
The ions are arranged by excitation potential. 
In the second column we propose a prototype with published spectrum.} 
\centering 
\begin{tabular}{l c c c c c c c c c c c c c } 
\hline
&&\multicolumn{12}{c}{ion/wavelength [\AA]}\\ [0.5ex]
\hline 
Spectral  & Example & \ion{H}{i}  &  \ion{He}{i}  &  \ion{C}{iii}  &  \ion{C}{iii}  &  \ion{He}{ii}  &  \ion{He}{ii}   &  \ion{C}{iv}  &  \ion{C}{iv}  &  \ion{N}{v}  &  \ion{N}{v}  &  \ion{O}{vi}  &  \ion{O}{vi}    \\ [0.5ex]
classif.  & & 4340 & 4471 & 4649 & 5696 & 4686 & 5412 & 5806 & 4650 & 4603 & 4945 & 3822 & 5290    \\ [0.5ex] 
\hline  
early O(H)  & NGC~5307  &  A  &  A  &  -  &  -  &  A  &  A   &  E  &  -  &  A  &  -  &  E? &  E   \\
late O(H)   & Cn~3$-$1  &  A  &  A  &  A  &  -  &  A  &  A   &  A  &  -  &  -  &  -  &  -  &  -  \\
early Of(H) & M~1$-$53  &  A  &  A  &  E  &  -  &  E  &  A   &  E  &  -  &  A  &  -  &  ?  &  E  \\
late Of(H)  & IC~4593   &  A  &  A  &  A  &  -  &  E  &  A   &  ?  &  -  &  -  &  -  &  -  &  ?   \\
Of-WR(H)    & NGC~6543  &  A  &  ?  &  -  &  -  &  E$^{1}$  &  E   &  E  &  -  &  E  &  -  &  -  &  E   \\
O(He)       & He~2$-$64 &  -  &  -  &  -  &  -  &  A  &  A   &  -  &  -  &  E  &  E  &  -  &  -  \\
PG~1159     & Jn~1      &  -  &  -  &  -  &  -  &  A  &  A   &  E  &  A  &  -  &  -  &  E  &  E  \\
\hline \hline 
$[$WO$]$    & NGC~2867  &  -  &  -  &  -  &  -  &  E  &  E   &  E  &  E  &  -  &  -  &  E  &  E  \\
$[$WC$]$    & NGC~40    &  -  &  -  &  E  &  E  &  E  &  E   &  E  &  -  &  -  &  -  &  -  &  -   \\
$[$WN$]$    & IC~4663   &  -  &  -  &  -  &  -  &  E  &  E   &  -  &  -  &  E  &  E  &  -  &  E  \\
DA          & DeHt~5    &  A  &  -  &  -  &  -  &  -  &  -   &  -  &  -  &  -  &  -  &  -  &  -   \\
DAO         & HDW~3     &  A  &  -  &  -  &  -  &  A  &  -   &  -  &  -  &  -  &  -  &  -  &  -   \\
DO          & KPD$^{2}$ &  -  &  A  &  -  &  -  &  A  &  A   &  -  &  -  &  -  &  -  &  -  &  -   \\
\hline \hline 
\end{tabular}
\tablefoot{\\ $^{1}$ Wide line. $^{2}$ KPD~0005$+$5106}
\label{lineSPT} 
\end{table*}

%%%%%%%%%%%%%%%%%%%%%%%%%%%%%%%%%%%%%%%%%%%%%%%%%%%%%%%%%%%%%%%%%%%%%%%%%
%%%%%%%%%%%%%%%%%%%%%%%%%%%%%%%%%%%%%%%%%%%%%%%%%%%%%%%%%%%%%%%%%%%%%%%%%
%%%%%%%%%%%%%%%%%%%%%%%%%%%%%%%%%%%%%%%%%%%%%%%%%%%%%%%%%%%%%%%%%%%%%%%%%

\section{Review of stellar evolution}\label{revestE}

The main parameter that defines the stellar evolution is the mass.
It is commonly accepted that the mass range of a main sequence star 
that gives birth to a PN is 1.0 M$_{\odot}$ $\lesssim$ M$_{*}$ $\lesssim$ 8.0 M$_{\odot}$. 
However, this mass range is not a strict limit.

Firstly, the minimum mass required to form a star is 0.07 M$_{\odot}$, 
the hydrogen burning mass limit \citep{2000ARA&A..38..337C}.
Therefore, stars with mass in the range between 
0.07 M$_{\odot}$$\lesssim$M$_{*}$$\lesssim$1.0 M$_{\odot}$ 
do not evolve to a PN.
In fact, these stars did not have enough time to 
leave the main sequence \citep{2019IAUS..343...36M}.
\cite{2016ApJ...832...48S} estimated the globular cluster turn-off mass 
in about 0.88 M$_{\odot}$ for NGC~6624. In this sense, 
it should not have progenitor CSPN with masses lower than 0.88 M$_{\odot}$.

A PN is obtained when the gas expelled in the AGB phase is 
ionized by the remaining object at its core.
This requires a star with a certain minimum progenitor mass M$_\chi$. 
Objects with mass less than M$_\chi$ will not result in a nebula.
Observational data must be used to obtain M$_\chi$. 
Globular clusters are the oldest objects in the galaxy and the 
lifetime of a PN is about 20\,000 years. 
Hence, if it is possible to determine the progenitor mass of a 
PN that is physically linked to a globular cluster, this will 
be a good estimation for $M_\chi$. \cite{2017ApJ...836...93J} 
showed that the progenitor mass of the PN JaFu~1 
(planetary nebulae linked to the globular cluster Pal~6) is 0.8 M$_{\odot}$.

On the other hand, finding observational evidence for the upper limit on the stellar mass, 
required to originate a PN, is not so straightforward.
An option is to determine the mass of the progenitor star of a PN
linked to an open cluster.
\cite{2019NatAs.tmp..357F} showed that the progenitor mass
of the planetary nebula BMP~J1613$-$5406 
(object linked to the open cluster NGC~6067) is 5 M$_{\odot}$.

\begin{figure}
    \centering
    \includegraphics[width=0.9\columnwidth]{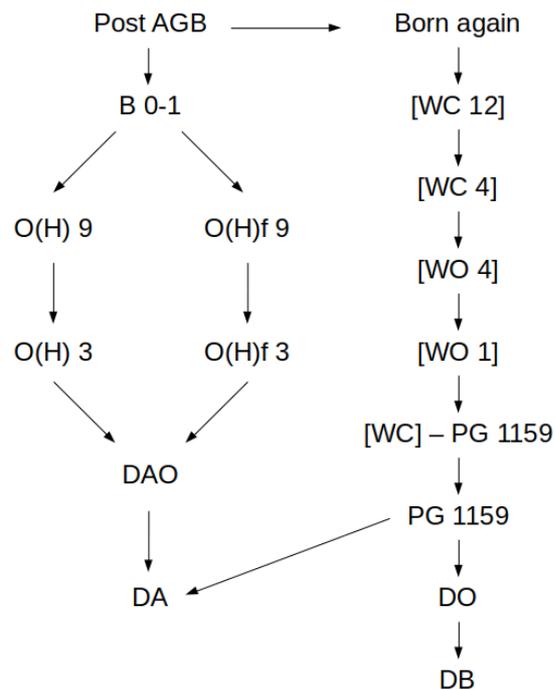}
    \caption{An evolutionary sequence model,
    the classical picture.}
    \label{fig:diag_evol}
\end{figure}

Now, how is the evolutionary sequence for a main sequence 
star with mass between 0.8 $-$ 5.0 M$_{\odot}$?
Which is the spectral-type sequence it passes through?
These questions still have no single answer.
\cite{1991ApJS...76...55I} developed a convincing theoretical model,
but it does not contemplate the spectral types.
Perhaps the sequence shown in Fig.~\ref{fig:diag_evol} is the most
classical picture
%w
\citep[see, e.g.,][]{1998A&A...330..265L,1994A&A...290..228Z,2001A&A...367..983P,2001Ap&SS.275...41K,2006PASP..118..183W}.
In addition, a more complex and complete (including several speculations) 
evolutionary path is presented by \cite{2014PhDT........76D}.
The stellar post-AGB evolution divides into two major channels of 
H-rich and H-poor stars \citep{2019MNRAS.489.1054L}. 
However, other evolutionary pathways exist \citep[see, e.g. Fig.~14 of][]{2014MNRAS.440.1345F}.
Nevertheless, it is necessary to clarify that these sequences are valid only for a single star.

The CSPNe with spectral type sdO, B[e], 
 O(He), [WN], hybrid  and Of$-$WR(H)
are not included in this classical evolutionary picture.

The origin of the sdO is yet not clear. Probably, this object comes 
from low mass progenitors \citep{2014A&A...565A..40R}, 
and could end up as a WD \citep{2013A&A...552A..25A}.
\cite{2015MNRAS.446..317A} observed that the sample of PNe with a sdO 
nucleus are faint and can be found at relatively high Galactic latitudes, 
which would suggest they are in a moderate or  evolved 
stage and evolving from low-mass progenitors. 
In the work of \cite{2009ARA&A..47..211H}, several evolutionary paths are considered.

The evolutionary state of B[e] CSPNe is unknown too.
The nebulae that contain these stars are compact and with high-density.
These nebulae are often classified as protoplanetary nebulae (PPNe) 
or young PNe \citep{1989IAUS..131..443C}. 
These objects have stellar hydrogen emissions and are therefore 
considered to belong to the H-rich group.
The O(He) stars are even more complicated,
\citep[see, e.g.,][]{2014A&A...566A.116R,1998A&A...338..651R}
as they are speculated to have multiple possible evolutionary paths.
One possible evolutionary path is 
Merger(CO WD + He WD)$\rightarrow$ RCrB $\rightarrow$ EHe $\rightarrow$ O(He) $\rightarrow$ DO 
\citep{2014A&A...566A.116R}.
It is necessary  to identify more objects of this type in order to clarify this point.

The same situation happens with the [WN] type. There are only three well-known objects, 
with an unclear evolutionary state. One possibility is 
that [WN] $\rightarrow$ O(He) \citep{2017IAUS..323..174T}.

The hybrid objects (or hybrid PG~1159-type) are strange objects \citep{2019MNRAS.489.1054L}.
This type include four objects, being one of them not in the core of a PN.
\cite{1996A&A...309..820D} suggest two possibilities:
[WC 12] $\rightarrow$ hybrid $\rightarrow$ PG~1159 or
[WC 12] $\rightarrow$ hybrid $\rightarrow$ DAO.

Finally, the Of$-$WR(H) CSPNe belong to the H-rich group.
There are not references regarding its evolutionary status. 
They may be binary objects.

%%%%%%%%%%%%%%%%%%%%%%%%%%%%%%%%%%%%%%%%%%%%%%%%%%%%%%%%%%%%%%%%%%%%%%%%%
%%%%%%%%%%%%%%%%%%%%%%%%%%%%%%%%%%%%%%%%%%%%%%%%%%%%%%%%%%%%%%%%%%%%%%%%%
%%%%%%%%%%%%%%%%%%%%%%%%%%%%%%%%%%%%%%%%%%%%%%%%%%%%%%%%%%%%%%%%%%%%%%%%%

\section{The catalogue of CSPNe}\label{catal}

This new version of the catalogue contains 620 galactic objects, 
representing an increase of 25\% compared to the previous version. 
However, it is not only in the number of objects that the real impact of the new version lies. 
We also report many spectral type changes. For example,
the core of He~2$-$47 that was previously classified as a [WC], 
has now been catalogued as an O(H)-type star based on newer, better quality spectra.
Besides, new high-quality spectra made possible to improve and 
refine the previous classification, as is the case of Sa~1$-$8 
(from OB type to O(H)4$-$8~III). 
In addition, some objects were rejected in the new version 
because we currently know that they are not PNe.
This is the case, for example, of Sh~2$-$128
\citep{2003AJ....126.1861B}.
Moreover, in this new version, we include  physical parameters such as temperature, 
luminosity, and surface gravity, which were obtained from the available literature. 
On the other hand, we added the apparent magnitude 
(a key parameter for planning an observation), and
a bibliography reference where the stellar spectrum is shown.
This allows future researchers to re-evaluate the spectral 
classification or to obtain a better quality spectrum.
In this sense, 
it is important to note that there are few spectra available 
in the literature, in particular of [WR]-type.
Also, in general, the spectral classification of these objects 
has not been confronted with new observations.
\cite{1998MNRAS.296..367C} and
\cite{2003A&A...403..659A} are perhaps the two most important works dealing with spectral
classification of [WR].
Both works include important object samples
however the published spectra do not
allow to properly check the spectral classification.
The catalogue is divided into two tables.
The former with the spectral type and the second one with  physical parameters. 
In the following sections, we describe each column of the two main tables.

%%%%%%%%%%%%%%%%%%%%%%%%%%%%%%%%%%%%%%%%%%%%%%%%%%%%%%%%%%%
%%%%%%%%%%%%%%%%%%%%%%%%%%%%%%%%%%%%%%%%%%%%%%%%%%%%%%%%%%%
%%%%%%%%%%%%%%%%%%%%%%%%%%%%%%%%%%%%%%%%%%%%%%%%%%%%%%%%%%%

\subsection{Table~\ref{tabla1}: spectral types}\label{des-1}

\begin{itemize}
\item c1:
The PN~G designation, taken from SECGPN.
\item c2:
The flag indicates that the object could not be a PN.
\item c3:
The common name of the object, taken from SECGPN and MASH.
\item c4-5:
The equatorial coordinates (J2000.0).
Frequently, it is not clear which is the CSPN
either because the star is not at the geometric center of the nebula, 
due to the interaction with the ISM or because of the object lies
in crowded star fields.
For example the PN Sh~2$-$71
\citep{2015MNRAS.451..870M} or K~4$-$37 \citep{2017MNRAS.466.2151M}.
The problem worsens when the PN has a large angular size, 
which may lead to incorrect identification of the CSPN
\citep[e.g.][]{2009ApJ...707L..32S, 1988AJ.....96..997K}.
A couple of articles that may be useful in identifying 
the position of a central star are
 \cite{2003A&A...408.1029K} and \cite{2020arXiv200108266C}.
\item c6-9:
Spectral classification (SpT) of the CSPNe and reference
(t.w. means this work).
We use the most current SpTs, unless there is a large discrepancy.
In that case, we include both classifications.
All of them are described in Section~\ref{st}.
We maintain the notation given by different authors,
in this context we keep the classification wels.
Nevertheless, the spectrum can change in a short period of time.
 \cite{2014AJ....148...44B} showed that the CSPN of Lo~4
 had a short-time mass loss episode, 
 manifested by the appearance of new emission lines.
 \item c10:
 Reference of multiplicity, taken from the  actual web
 of Dr. David Jones\footnote{http://www.drdjones.net/bCSPN/}.
 We do not included binaries detected by J or I-band excess.
\item c11: Reference of the optical spectra.
\end{itemize}

%%%%%%%%%%%%%%%%%%%%%%%%%%%%%%%%%%%%%%%%%%%%%%%%%%%%%%%%%%%
%%%%%%%%%%%%%%%%%%%%%%%%%%%%%%%%%%%%%%%%%%%%%%%%%%%%%%%%%%%
%%%%%%%%%%%%%%%%%%%%%%%%%%%%%%%%%%%%%%%%%%%%%%%%%%%%%%%%%%%

\subsection{Table~\ref{tabla2}: physical parameters}\label{des-2}

\begin{itemize}
\item c1:
PN~G designation. The first nine objects do not have PN~G, then we use the
\textit{recno} number of Table~\ref{tabla1}.
\item c2-3: Surface gravity ($log~g$, cm s$^{-2}$) and reference.
t.w. means this work.
We computed $log~g$ according to equation 17
of \cite{1993ApJS...88..137Z}
with data take from \cite{2007A&A...467L..29G}.
\item c4: The method by which the temperature was determined.
We prefer the effective temperature (T$_{eff}$).
Nevertheless, in the catalogue we include other temperatures,
Zanstra temperatures T$_Z$(He~I) or T$_Z$(He~II)
\citep{2003MNRAS.344..501P}
or  the photoionization (equivalent black-body) temperature
\citep{2006A&A...451..925G}.
\cite{1995MNRAS.273..812H} show the different results obtained for 
CSPN temperature using different methods.
\item c5-6: Temperature and reference.
In the case of the H-rich group, it is possible to
estimate the temperature inspecting the spectra 
and looking for \ion{He}{i} absorption lines. 
The absence of these lines indicates that the central star is hotter than 70 k
\citep{2016MNRAS.455.3413K}.
\item c7-8: Bolometric luminosity 
($\log (L_\star/L_{\odot}$))
and reference
 \citep[e.g.][]{1993ApJS...88..137Z}.
\item c9-10: The apparent magnitude and reference.
We indicate the band in which the magnitude is reported. 
When available, the visual magnitude was preferred.
For those objects catalogued as photometric variables,
we report an average magnitude. For binary systems, it 
is probable that the reported magnitude corresponds to the bright source,
 which is not necessary the ionizing source. Moreover, when the CSPN is binary, 
 it is usual that it presents photometric variations.
\end{itemize}

The physical parameters that we report in the tables are 
the most up-to-date and post 1982.
In addition, we include the uncertainty, 
only in cases where the original author published them.
Nevertheless,
there are objects that undergo a fast evolution of their physical parameters.
For example He~3$-$1357 \citep{2014A&A...565A..40R}.
In this sense, 
the physical parameters of the CSPN have importance from a statistical viewpoint.

%%%%%%%%%%%%%%%%%%%%%%%%%%%%%%%%%%%%%%%%%%%%%%%%%%%%%%%%%%%%%%%%%%%%%%%%%
%%%%%%%%%%%%%%%%%%%%%%%%%%%%%%%%%%%%%%%%%%%%%%%%%%%%%%%%%%%%%%%%%%%%%%%%%
%%%%%%%%%%%%%%%%%%%%%%%%%%%%%%%%%%%%%%%%%%%%%%%%%%%%%%%%%%%%%%%%%%%%%%%%%

\subsection{Multiplicity vs. spectral type}

CSPNe confirmed to be part of a binary system are described in
Table~\ref{tabla1} (column~10).
It is known that many stars are binary, but in few cases it is possible 
to have the spectrum of the two stars separately. 
The criterion for describing the spectral type in these cases is
$ SpT_1+SpT_2 $.
Where the first star is the brightest,
not necessarily the ionizing star of the nebula.
In the case in which the spectral type of a star is not known, 
it is indicated as $SpT_1+?$.

The fact that recently the number of CSPNe that are binary 
systems have been increased considerably, means 
that we have to rethink which component would be considered the CSPN.
An appropriate response would be: the star of the binary 
system responsible for the ionization of the nebula.
Nevertheless, it is possible that both stars of the binary system have 
enough  temperature to ionize the nebula.
This is the case of He~2$-$428, where
the first star has T$_{eff}$=48$\pm$7 kk and 
the second star has T$_{eff}$=46$\pm$7 kk
\citep{2018Galax...6...88R}.

Two objects require clarification
because there are no references reporting its binarity:
i) Cn~1$-$1: there is evidence indicating that this object is a D'-type
symbiotic star \citep{1995PASP..107..462G}. 
ii) PHR~J0905$-$4753: in the MASH catalogue the central star is identified 
with a spectral type A8. If this is correct,
then it is indeed a binary system.

%%%%%%%%%%%%%%%%%%%%%%%%%%%%%%%%%%%%%%%%%%%%%%%%%%%%%%%%%%%%%%%%%%%%%%%%%
%%%%%%%%%%%%%%%%%%%%%%%%%%%%%%%%%%%%%%%%%%%%%%%%%%%%%%%%%%%%%%%%%%%%%%%%%
%%%%%%%%%%%%%%%%%%%%%%%%%%%%%%%%%%%%%%%%%%%%%%%%%%%%%%%%%%%%%%%%%%%%%%%%%

\section{Results}\label{reul}

\subsection{Dichotomy between H rich and poor CSPNe}\label{dichoto}

We separate the CSPNe of our sample according to their atmospheric 
hydrogen abundance into H-rich and H-poor (see Table~\ref{tab1}).
The ratio H-rich/H-poor in the present catalogue is 2:1,
which is greater than the one found by
\cite{2011A&A...526A...6W}. %3.0
Nevertheless, it is 
still far from the value determined by
\cite{1991IAUS..145..375M}.
The large number of objects included in our present catalogue 
reinforce the idea that the H-poor population is larger than previously thought.

\begin{figure}
    \centering
    \includegraphics[width=0.9\columnwidth]{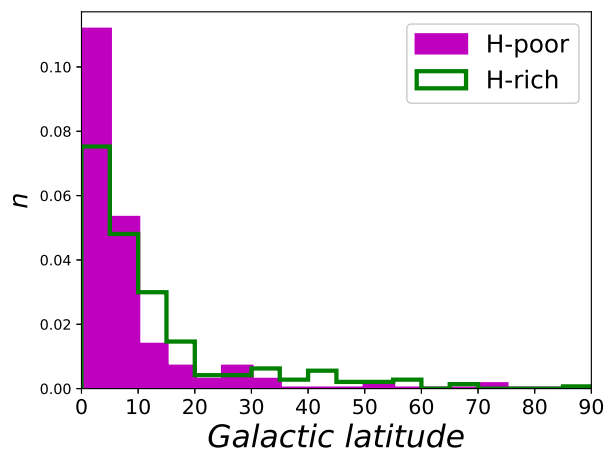}
    \caption{Normalized distributions in Galactic latitude of CSPNe
    (of confirmed and possible PNe) that belong to H-rich 
    and H-poor group. Each bin has a width of 5~deg.}
    \label{fig-histo}
\end{figure}

Applying the Kolmogorov-Smirnov test 
to compare the Galactic latitude distributions of H-rich and H-poor populations, we find 
D=0.2 and a p-value=0.00049.
 This is in agreement with the results found by \cite{2011A&A...526A...6W}, 
 which implies that the distribution in Galactic latitudes of H-rich 
 and H-poor stars are different (Fig.~\ref{fig-histo}).
This is an indication that the progenitors masses and ages of both populations are different. 
The progenitors of H-deficient stars expected to be more massive and 
younger than those of their H-rich counterparts. 
It would be interesting to analyse whether this can be related to 
the observed dearth of H-deficient white dwarfs in old globular clusters 
\citep{2004A&A...420..515M, 2009ApJ...705..398D}, 
as reasons behind this feature are still not clear \citep{2018ApJ...867...62W}.

The physical parameters that we have collected from the
literature allow us to characterize the CSPNe population
(Table~\ref{param}).
 Although it is generally accepted that hydrogen requires a minimum temperature 
of 10\,000 K to ionize,
the minimum temperature of a CSPN is 13\,000 K. This value agrees with what was expected.

\begin{figure}
    \centering
    \includegraphics[width=0.9\columnwidth]{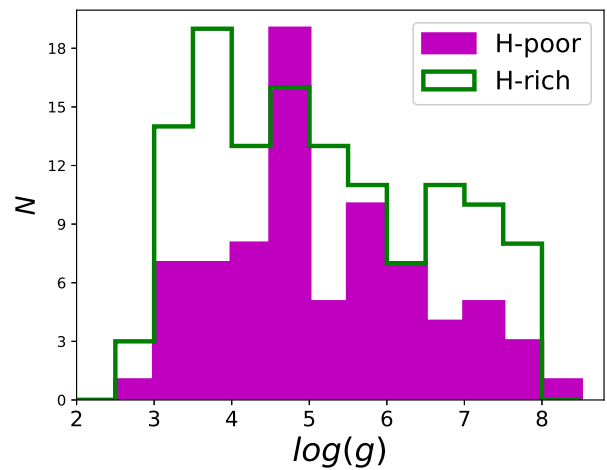}
    \caption{Distribution in surface gravity of CSPNe. }
    \label{histo-g}
\end{figure}

\begin{figure}
    \centering
    \includegraphics[width=0.9\columnwidth]{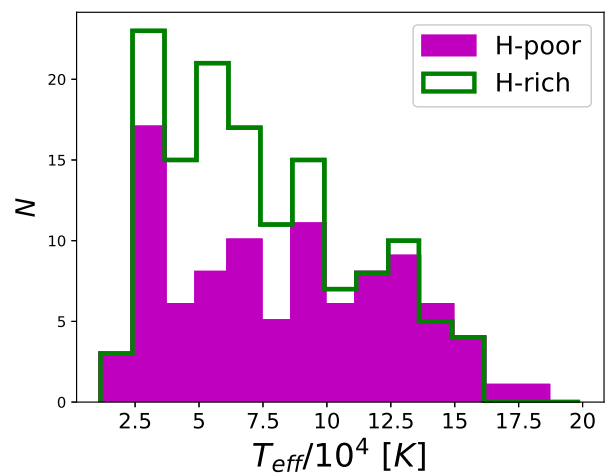}
    \caption{Distribution in temperature of CSPNe.}
    \label{histo-Teff}
\end{figure}

\begin{figure}
    \centering
    \includegraphics[width=0.9\columnwidth]{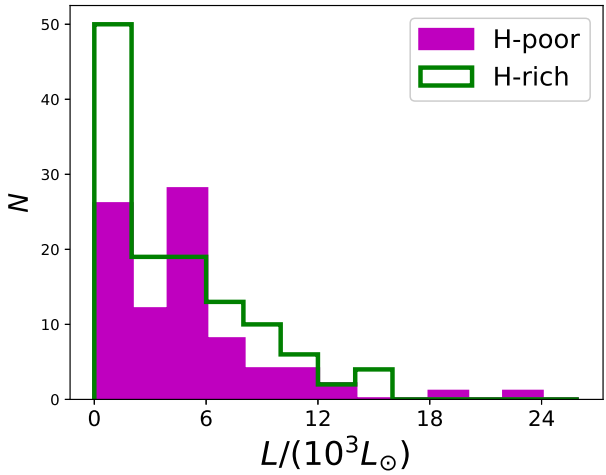}
    \caption{Distribution in luminosity of CSPNe. }
    \label{histo-l}
\end{figure}

According to the values presented in Table~\ref{param} 
(Figs.~\ref{histo-g}, \ref{histo-Teff}, and \ref{histo-l}), 
the H-poor population appears to have grater surface gravity than 
the H-rich one, except for the most evolved objects. 
With regard to the effective temperature, the H-poor group is clearly hotter.
Luminosity shows a more complex situation. 
Still a trend can be appreciated in which more evolved H-poor objects display a higher luminosity.

\begin{figure}
    \centering
    \includegraphics[width=0.9\columnwidth]{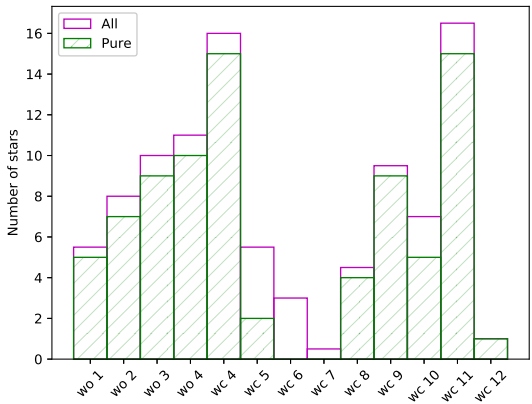}
    \caption{Distributions of the [WR] type CSPN over the subtype. 
The sample "Pure" exclude objects with imprecise classification
(e.g. [WC 5-6]). The sample "All" includes these objects with the criteria: 
an object with classification [WC 5-6], adds 0.5 to the frequency of [WC 5] 
and adds 0.5 to the frequency of [WC 6].}
    \label{sub-wr}
\end{figure}

Moreover, Fig.~\ref{sub-wr} shows the [WR]
subtype distribution, where a 
dearth of objects with subtypes 5-7 is apparent. 
This feature is seen both in our restricted sample with definite spectral types 
(our "pure" sample), as well as in the complete sample 
that includes also uncertain spectral types. 
This result is consistent with previous results in the literature
 \cite{2009PhDT.......166T}. 
 Even more, given that our sample is significantly larger
 than that discussed by \cite{2009PhDT.......166T}, we consider this 
 result as a confirmation that the lack of subtypes 5-7 in [WR]-CSPNe is a real feature. 
 In principle this gap could be explained in two different ways. 
 On the one hand it could be the consequence of faster evolutionary speeds at that 
 particular stage and, on the other, it could be the consequence that  
 specific photospheric conditions are 
 necessary for [WR]-CSPNe to be classified as [WC 5-7]. 
 Interestingly, a concomitant gap is not apparent in the sample of H-deficient CSPNe 
 with derived effective temperatures, suggesting that the latter might be the right explanation. 
 This hypothesis will need to be teste by specifically tailored future studies.
 
 Regarding Of-type CSPNe, within these objects  early subtypes are 
 predominant, with  the coolest object being Hen~2$-$138,
 classified as O(H)7-9~f. \cite{1990A&A...229..152M} proposes that O stars are less luminous than the Of stars. 
 However, we do not find a significant difference between both groups. 
 We find that O-type CSPNe have  $\overline{ (L_\star/L_{\odot})}$=5.3$\times 10^3$ and $\overline{ (L_\star/L_{\odot})}$=5.7 $\times 10^3$ for Of-type.

\begin{figure}
    \centering
    \includegraphics[width=0.9\columnwidth]{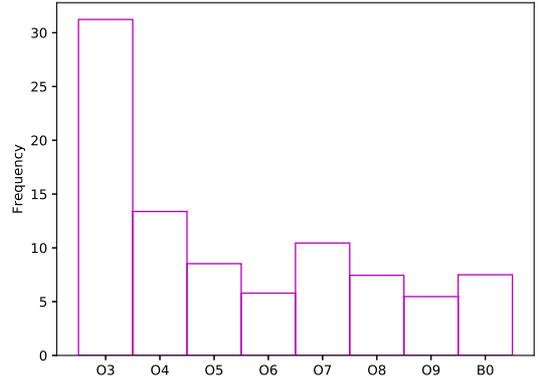}
    \caption{Distributions of the O-type CSPN over the subtype. 
    Criterion used for objects with dubious classifications (e.g. O5-9) 
    is the same as that described in Fig.~\ref{sub-wr}.}
    \label{sub-ott}
\end{figure}

A peculiar situation can be observed by comparing 
the number of objects 
with different spectral types.
The number of objects classified as [WC 4-12]
is larger than [WO 1-4]. 
We interpret this as a difference in evolutionary speeds.
 i.e., these objects spend more time in its earliest evolutionary stage than in the subsequent ones.
Interestingly, this is in clear contradiction with what is expected 
from the late thermal pulse scenario. 
Stellar evolution models 
are predicted to slow their evolution as they 
reach their maximum effective temperature \cite[i.e. the "knee" of 
the evolutionary tracks in the HR-diagram, see the supplementary material in  ][]{2018NatAs...2..784G}. 
Alternatively, this could be a consequence of the larger luminosity 
expected during the earlier stellar evolution, which leads to a larger 
volume in which these stars can be detected. 
Volume limited samples are required to solve this issue.
On the other hand, the distributions of the O-type CSPNe
over the subtypes is different
(Fig.~\ref{sub-ott}).
This distribution is in general flat, but with a remarkable number of objects classified as O3.
This may be due to a selection effect.
Since for the O-type CSPNe the spectral classification is more qualitative that in case of the $[$WR$]$, 
many objects have doubtful classifications. 
But, in the case of the O3 CSPNe, these are easily identifiable by the absorption of the \ion{N}{v}.

Nevertheless, this might be due to the saturation of a qualitative 
classification scheme, originally constructed for main sequence object, 
when applied to the much hotter CSPNe.

\begin{figure}
    \centering
    \includegraphics[width=0.9\columnwidth]{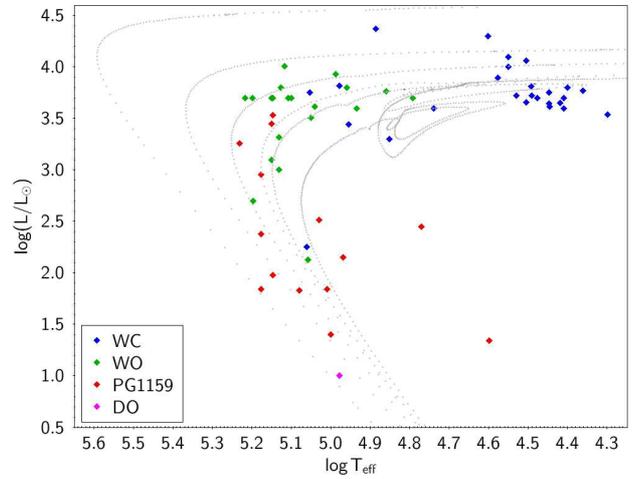}
    \caption{HR diagram of H-poor CSPNe, according subtype. H-deficient tracks correspond to the post-VLTP sequences computed by  \cite{2006A&A...454..845M}, which display He, C, and O photospheric abundances similar to those of  [WR]-CSPNe and PG~1159 stars (M$_{CSPN}$=0.515, 0.542, 0.584, 0.664 y 0.870 $M_\odot$, from right to left).}
    \label{hr-s-p}
\end{figure}

\begin{figure}
    \centering
    \includegraphics[width=0.9\columnwidth]{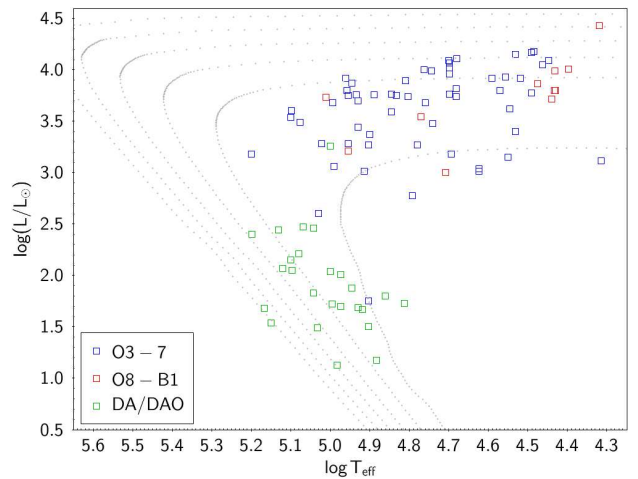}
    \caption{HR diagram of H-rich CSPNe, according subtype. H-rich track  
    are described in Section~\ref{hr}.}
    \label{hr-s-r}
\end{figure}

Finally, Figs.~\ref{hr-s-p} and \ref{hr-s-r} display the distribution 
of spectral subtypes for both the H-poor and H-rich populations. 
According to the evolutionary sequence described in Fig.~\ref{fig:diag_evol}, 
stars are thought to evolve along evolutionary tracks, first increasing 
their effective temperature and then becoming dimmer and entering the white dwarf cooling sequence.

\begin{table}
\caption[]{Summary of the spectral types of CSPNe compiled
in our catalogue, grouped by their atmospheric hydrogen abundance.
We include confirmed and probable spectral types.
Here, we have discarded 6 objects without any specific spectral type.}
\label{tab1}
\centering
\setlength{\tabcolsep}{0.2mm}
\begin{tabular}{lc | lc | lc}
\hline%\noalign{\smallskip}
% & & & & & \\[-1em]
\multicolumn{2}{c}{H-rich}    &     \multicolumn{2}{|c|}{}     &  \multicolumn{2}{c}{H-poor}  \\
 SpT             & sample       & SpT  & $ $ $ $ sample & SpT  &  sample\\
%\noalign{\smallskip}
\hline%\noalign{\smallskip}
 & & & & & \\[-0.7em]
O3-B1       & 133 & cont.       & 22 & [WC 4-12]    & 68 \\
Of          & 31  & hybrid      & 3  & [WO 1-4]     & 37 \\
Of-WR(H)    & 3   & SySt        & 11 & [WN]         & 8 \\
B2-M9       & 40  & Blue        & 50 & [WR]         & 10 \\
DA, WD      & 14  & EL          &  8 & PG~1159      & 17 \\
DAO         & 28  & wels        & 77 & [WC]-PG~1159 & 2 \\
sdO         & 12  &             &    & O(He)        & 5 \\
hgO(H)      & 12  &             &    & O(C)         & 1 \\
B[e]        & 6   &             &    & DO           & 4 \\
H-rich      & 11  &             &    & H-poor       & 1  \\
\hline
 & & & & & \\[-0.7em]
%\noalign{\smallskip}%
Total         & 290 & Total         & 171 & Total      & 153 \\
  \hline
  \hline
\end{tabular}
\tablefoot{\\ $^{1}$The O3-B1 group include objects classified as O(H), O and OB. \\ $^{2}$ [WR] group include the two objects classified as [WN/C] and [WO]-[WC 8].}
\end{table}

\begin{table}[h!]
  \begin{center}
    \caption{Average physical parameter of our catalogue.
    We only included objects with accurate SpT,
    and exclude objects with confirmed binary nucleus.
    Between parentheses the number of objects is indicated.}
    \label{param}
    \begin{tabular}{l|c|c|c}
   
   Population & $\overline{log g}$ & $\overline{T_{eff}}$ $\times 10^3$ & $\overline{ (L_\star/L_{\odot})}$ $\times 10^3$ \\
      \hline
     $[$WC 4-12$]$   & 4.17 (33) & 50 (41) & 6.3 (36)  \\
     $[$WO 1-4$]$    & 5.37 (20) & 110 (23) & 4.2 (21)  \\
     PG~1159         & 6.68 (10) & 117 (14) & 0.7 (13)  \\
      \hline
     O8 - B0         & 3.51 (13) & 38 (16) & 10 (14)  \\
     O3-7            & 4.52 (26) & 71 (29) & 5.9 (29)  \\
     hgO(H), DA, DAO & 6.93 (31) & 99 (33) & 0.1 (25)  \\
           \hline \hline
    \end{tabular}
  \end{center}
\end{table}

%%%%%%%%%%%%%%%%%%%%%%%%%%%%%%%%%%%%%%%%%%%%%%%%%%%%%%%%%%%%%%%%%%%%%%%%%
%%%%%%%%%%%%%%%%%%%%%%%%%%%%%%%%%%%%%%%%%%%%%%%%%%%%%%%%%%%%%%%%%%%%%%%%%
%%%%%%%%%%%%%%%%%%%%%%%%%%%%%%%%%%%%%%%%%%%%%%%%%%%%%%%%%%%%%%%%%%%%%%%%%

\subsection{Binary CSPNe}\label{binoCSPN}

While there are continuous efforts focused on the search for new binary in PNe 
\citep[e.g.,][]{2015MNRAS.448.3132D}, the fraction of PNe with binary nuclei 
is still relatively small\footnote{A regularly updated catalogue of binary 
CSPNe is mantained by David Jones can be found at \url{http://www.drdjones.net/bcspn/.}}.
In our catalogue there are 117 PNe
with a confirmed binary system at its nuclei. 
We do not include the SySt stars.
Eleven of them belong to the H-poor group
([WR], DO, H-poor, O(He), PG~1159, and [WN]), plus an object classified as [WR]/wels (Vy~1$-$2).
The O, O(H), O(H)f, sdO, H-rich, DAO, and hgO(H) stars totalize 62 objects, together with nine objects
that do not have a specific spectral type.
In addition, there are 34 objects with
late spectral types.

For the vast majority of binary systems there are only spectroscopic 
data for the brightest star, so there are few systems in which we 
know the spectral type of the two stars of the system. 
No binary systems composed by stars with different H abundances in its atmosphere 
(i.e. H-rich + H-poor) are currently known. 
In this sense, and for statistical purposes, 
we adopt the criterion of using the H abundance
of the brightest star.

With the above clarifications,
we found that 10.3\% of confirmed binaries belong to the
H-poor group and 82.1\% to the
H-rich group.
Consequently, there is a pronounced tendency,
even more so considering that the H-rich ones are twice as many in the general population,  
for binary stars to occur more frequently in stars with hydrogen in their atmosphere. 

The lack of H-deficient central stars in binary systems is a natural 
expectation from the proposed scenarios for their formation, 
namely the merger and the late thermal pulse. 
This is particularly true given that close binary systems 
are much easier to be detected than wide binaries. 
Mass transfer in a close binary system is expected to prevent the 
natural occurrence of the TP-AGB phase, making the late thermal pulse scenario unlikely. 
And, while a merger after a common envelope phase is a possible 
outcome, if this merger leads to a H-deficient star it will be a 
single H-deficient star unless the original system was triple, and in a  specific configuration.

%%%%%%%%%%%%%%%%%%%%%%%%%%%%%%%%%%%%%%%%%%%%%%%%%%%%%%%%%%%%%%%%%%%%%%%%%
%%%%%%%%%%%%%%%%%%%%%%%%%%%%%%%%%%%%%%%%%%%%%%%%%%%%%%%%%%%%%%%%%%%%%%%%%
%%%%%%%%%%%%%%%%%%%%%%%%%%%%%%%%%%%%%%%%%%%%%%%%%%%%%%%%%%%%%%%%%%%%%%%%%

\subsection{Comparison with stellar evolution models. Stellar masses and ages}\label{hr}

\begin{figure}
    \centering
    \includegraphics[width=0.9\columnwidth]{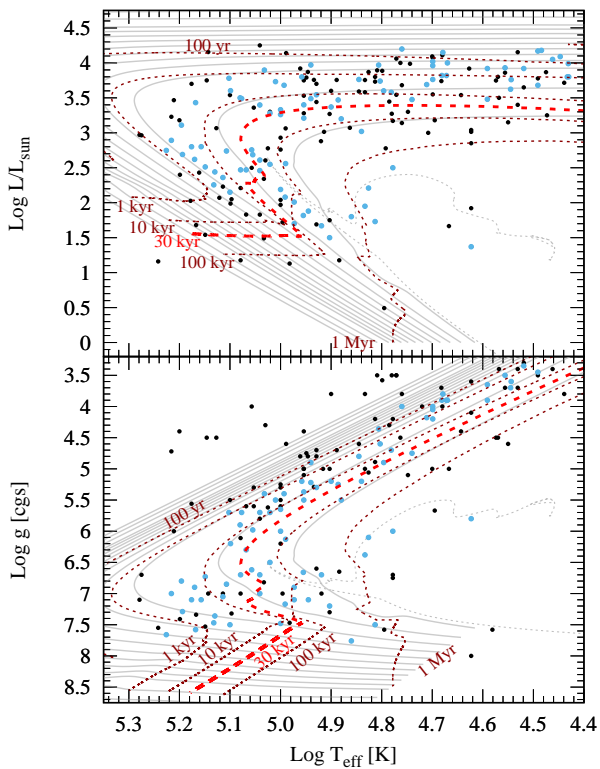}
    \caption{Location of the selected sample of H-rich stars 
    (213 objects in total) in both the HR (upper panel) and Kiel 
    (lower panel) diagrams. Light blue points symbols show the location of 
    the subsample for which interpolation in both diagrams indicate a 
    consistent age and mass (88 objects, see text). Grey continuous 
    lines show the location of interpolated tracks for different masses 
    (0.5, 0.55, 0.6, 0.65, 0.7, 0.75, 0.8, 0.85, 0.9, 0.95, 1, 1.1, 1.2, 1.3 $M_\odot$, 
    and $Z_0=0.01$, from right to left). Brown and red lines show isochroned 
    in the post-AGB evolution. The doted grey line indicates a typical post-EHB evolution.}
    \label{fig:HR_Kiel}
\end{figure}   
\begin{figure}
    \centering
    \includegraphics[width=0.9\columnwidth]{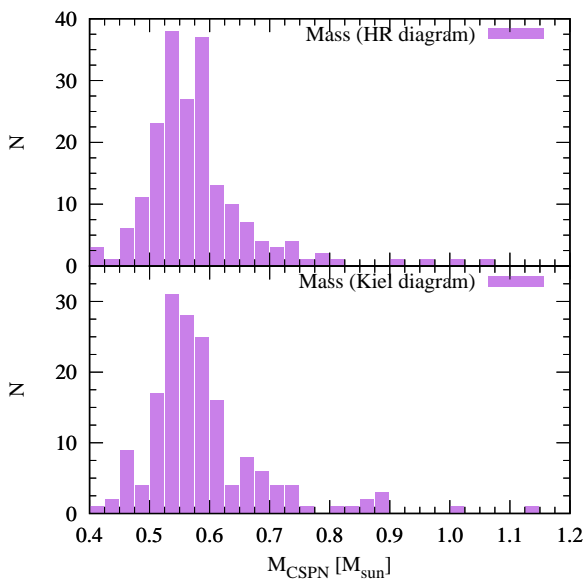}
    \caption{Masses derived from interpolation of the full H-rich sample 
    (213 objects, see Fig.~\ref{fig:HR_Kiel}) with  the $Z=0.01$ model grid of \cite{2016A&A...588A..25M}. 
    Upper panel: Masses derived from interpolation in the HR diagram. 
    Lower panel: Masses derived from interpolation in the Kiel diagram ($\log g$ - $\log T_{\rm eff}$).}
    \label{fig:full_mass}
\end{figure}   
\begin{figure}
    \centering
    \includegraphics[width=0.9\columnwidth]{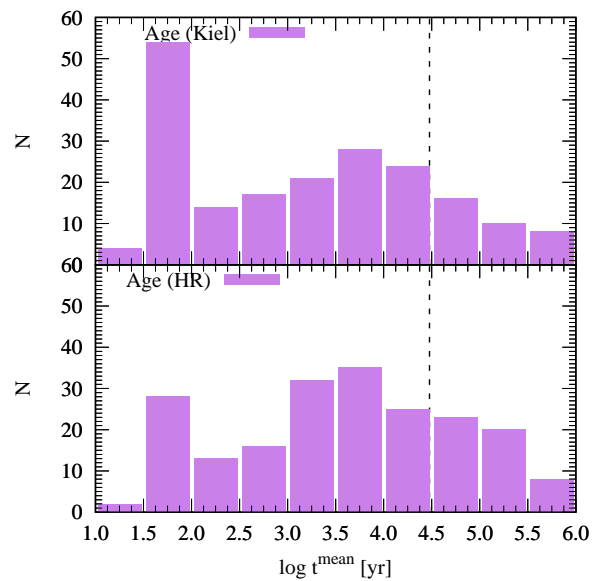}
    \caption{Ages derived from interpolation of the full H-rich sample 
    (213 objects, see Fig.~\ref{fig:HR_Kiel}) with  the $Z=0.01$ model 
    grid of \cite{2016A&A...588A..25M}. Upper panel: Masses derived 
    from interpolation in the HR diagram. Lower panel: Masses derived 
    from interpolation in the Kiel diagram ($\log g$ - $\log T_{\rm eff}$). 
    The vertical dashed line indicates the canonical value of 30 Kyr 
    value beyond which PNe are not expected to survive.}
    \label{fig:full_age}
\end{figure}

From the comparison of the surface $\log L_\star$, $\log g$, and $\log T_{\rm eff}$ 
values of our catalogue stars with those predicted by stellar evolution models 
\citep{2016A&A...588A..25M} it is possible to derive ages and masses for the CSPNe. 
Due to the absence of a grid of appropriate stellar evolution models for 
H-deficient stars, this can only be done for those with normal 
H/He surface compositions (i.e. those with solar-like H/He compositions). 
A comparison of 213 of our H-rich catalogue stars with the $Z=0.01$ models of 
\cite{2016A&A...588A..25M}  is shown in Fig.~\ref{fig:HR_Kiel} both for the 
 $\log L_\star$-$\log T_{\rm eff}$ (i.e. HR diagram) diagram 
 and the $\log g$-$\log T_{\rm eff}$ (i.e. Kiel diagram). 
 From the total of 213 objects only 175 have values of all three 
 surface parameters and can be plotted in both diagrams.
 In this sample we exclude doubtful PNe, SySt, hybrid and objects classified as "O?".
 Also there were excluded those objects where two contradictory spectral classifications are reported,
 i.e. H-rich and H-poor (e.g. IC~4776).
%se dejaron las clasificadas como "O" y las "cont." y las binarias.

  By interpolating the location of our sample stars in both the HR and Kiel 
  diagrams we can estimate their masses  
  (Fig.~\ref{fig:full_mass}) and ages (Fig.~\ref{fig:full_age}).
 A couple of things are particularly noteworthy in the full H-rich sample shown in 
 Fig.~\ref{fig:HR_Kiel}. 
 On the one hand the masses of CSPNe inferred from the Kiel and HR diagram interpolations 
 agree quite well with our current expectations from single post-AGB stellar evolution 
 (see histograms in Fig.~\ref{fig:full_mass}), with peaks around ($\sim0.55 M_\odot$). 
 Both diagrams show, 
 however, slightly different pictures, while interpolation in the HR diagram shows a double peak, 
 the Kiel diagram shows a wider peak between  $0.525 M_\odot$ and $0.600 M_\odot$. 
 On the other hand, some features are clearly at variance with expectation from stellar evolution models. 
 One is the group of objects at low surface gravities, well beyond the most massive model 
 (Fig.~\ref{fig:HR_Kiel}, lower panel). 
 Such object would have masses well beyond the Chandrasekhar mass if they 
 were post-AGB stars and should not be there. 
 In addition, there is a rather large number of stars with inferred masses 
 below $0.5 M_\odot$, something that is not expected from the 
 point of view of single stellar evolution. 
 Finally, interpolated/extrapolated ages and masses in the full H-rich 
 sample show a relatively large number of objects beyond the 30 Kyr isochrone, 
 i.e. well beyond the expected lifetime for true PNe, as well as a large number 
 of  young objects with interpolated ages below 100 yr (Fig.~\ref{fig:full_age}). 

\begin{figure}
    \centering
    \includegraphics[width=0.9\columnwidth]{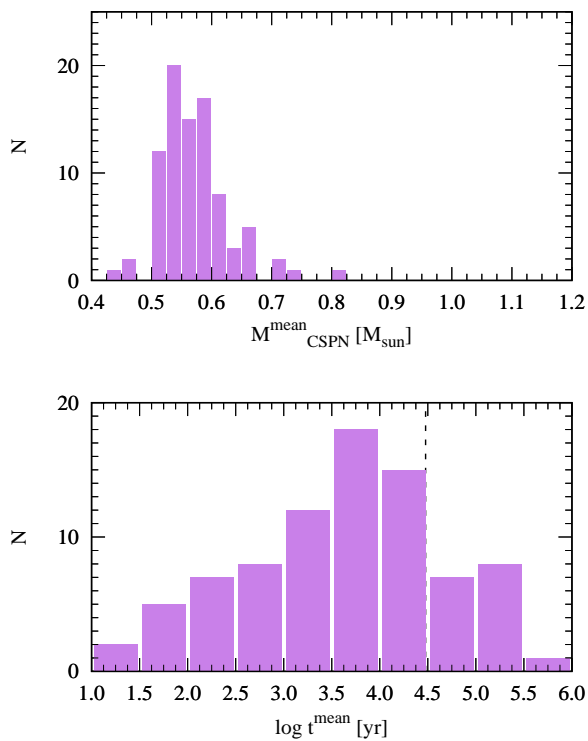}
    \caption{Mean values of the ages and masses for the reduced sample of 88 objects 
    (blue dots in Fig.~\ref{fig:HR_Kiel}) for which interpolation in the 
    Kiel and HR diagrams yield consistent results. 
    Age (mass) of each object is calculated as the geometric (arithmetic) mean 
    of the values yield by the interpolation in each diagram. 
    The vertical dashed line indicates the canonical value of 30 Kyr value 
    beyond which PNe are not expected to survive.}
    \label{fig:mean_mass_age}
\end{figure}

 To analyse which of the previous features are real and which ones 
 are artifacts due to poor determinations of the stellar parameters 
 of individual stars, we restricted to those stars that have independent 
 determinations of all three surface parameters ($\log L_\star$, $\log T_{\rm eff}$ and $\log g$).
 From these we selected the subsample for which both the HR and Kiel 
 diagram interpolations give consistent values of age and 
 mass\footnote{Which we arbitrarily set to be in agreement within an order 
 of magnitude for age and a difference of less than a 10\% for mass.}. 
 This smaller sample is shown as light blue dots in Fig.~\ref{fig:HR_Kiel}. 
 Consists of 88 objects out of the 175 objects (i.e. 50\%) 
 that have values of $\log L_\star$, $\log T_{\rm eff}$ and $\log g$. 
 The consistency of the values of $\log L_\star$, $\log T_{\rm eff}$ 
 and $\log g$ gives us confidence that no large errors are present in the 
 derived masses and ages. Fig.~\ref{fig:mean_mass_age} shows the masses and ages of this reduced sample. 
 Remarkably, many of the weird features seen in the full sample (Figs.~\ref{fig:full_mass} 
 and \ref{fig:full_age}) disappear when we restrict ourselves to our internally consistent sample. 
 Specifically, no  low gravity objects remain, 
 and the number of  young CSPNe is strongly reduced. 
 Also, the mass values of the smaller, and consistent, sample show a sharp 
 cut-off at $\sim0.5 M_\odot$  much consistent with the expectations from stellar evolution theory. 
 Similarly, the post-AGB age distribution of CSPNe in this smaller sample shows 
 a steady rise until the typical timescales of PNe (10 to 30 Kyr) and a clear drop afterwards.

Interestingly, we see that in this reduced sample a small group of objects with 
inferred  low masses persist 
($M_\star$<$0.5 M_\odot$). As such low masses, the objects cannot be post-AGB stars, 
and these extrapolated masses cannot be taken at face value but just as an indication 
of a different evolutionary origin. These low-mass objects correspond to 
the high-gravity/low-luminosity blue points in Fig.~\ref{fig:HR_Kiel}. 
These objects are in fact located where post extreme horizontal branch (EHB) stars (a.k.a. "AGB-manqu\'e" stars) 
should be \cite[e.g.][]{2019A&A...627A..34M}. 
This alternative evolutionary path is displayed by the dotted grey line in Fig.~\ref{fig:HR_Kiel}. 
If these objects are indeed post-EHB stars then their masses  should be 
close to $0.5 M_\odot$, and due to the long timescale of this evolutionary stage the 
"PN" around them is probably ionized interstellar gas that was not ejected by the star. 

We conclude that, when we restrict our sample to those stars for 
which $\log L_\star$, $\log T_{\rm eff}$ and $\log g$ yield a 
consistent picture, the derived masses and ages are in  good 
agreement with the expectations of single star post-AGB evolution theory. 
We conclude that a 50\% of our sample with values of $\log L_\star$, $\log T_{\rm eff}$ 
and $\log g$ (88 out of 175 objects) show masses and ages consistent with single stellar evolution models. 
Although some of this objects seem to have derived ages too large to 
host a PN, we believe that in view of the ongoing debate of whether single 
stars can form PNe \citep{2017NatAs...1E.117J, 2019ibfe.book.....B}, this is a remarkable result. 
 This should not be taken as a claim that multiple systems do not form PNe or that companions do not play any role in the shaping of the PNe. But rather a hint that stars in isolation or in wide binaries, are able to form PNe as well.
As mentioned before, common envelope events are the reason for the formation of PNe with close binary central stars \citep{2018Galax...6...75I, 2019ibfe.book.....B}. 
There have been suggestions that substellar companions \citep{ 2018MNRAS.473..286S} or triple systems \citep{2016MNRAS.455.1584S} might be involved in the formation or shaping of non-spherical PNe.

%%%%%%%%%%%%%%%%%%%%%%%%%%%%%%%%%%%%%%%%%%%%%%%%%%%%%%%%%%%%%%%%%%%%%%%%%
%%%%%%%%%%%%%%%%%%%%%%%%%%%%%%%%%%%%%%%%%%%%%%%%%%%%%%%%%%%%%%%%%%%%%%%%%
%%%%%%%%%%%%%%%%%%%%%%%%%%%%%%%%%%%%%%%%%%%%%%%%%%%%%%%%%%%%%%%%%%%%%%%%%

\section{Summary and conclusions}
\label{conclu}

In this paper we present a new CSPNe catalogue comprising 620 stars that 
could be spectrally classified in a reliable way. 
This catalogue represents an expansion of the first version 
\citep{2011A&A...526A...6W} with the inclusion of 128 new objects. 
In addition to updating all available information, this version 
includes new physical parameters for the CSPNe, as well as the visual 
magnitude and the corresponding bibliographic reference 
of the spectra used for spectral classification. 
The information collected and processed in this catalogue represents 
an essential contribution both in theoretical studies and in the planning 
of future observations. 
For all this we understand that it constitutes 
a fundamental tool both for the field of planetary nebula research 
in particular, and for stellar evolution studies in general.

In relation to the completeness of the information included in this catalogue, of the total number of objects:
56.3\% have $log~g$,
69.8\% have $log~T$,
60.5\% have $log~L_\star$, and
74.4\% have magnitude.

It is important to note that the objects reported in this catalogue 
represent less than 20\% of the total PNe known to date. Furthermore, 
almost 30\% of these objects do not have any specific physical parameter determined. 
The origin of this lack of information is probably related to the 
low brightness of these objects, if we consider that 50\% of 
the sample has a visual magnitude lower than 17. 
This reveals an important information gap in this area of research, 
a gap that can only be overcome with new large-scale studies using 8-10 m class telescopes.
\\

The main conclusions can be summarized as follows:

\begin{itemize}

    \item[$-$] When we separate the CSPNe between H-rich and H-poor we find that the ratio H-rich/H-poor is 2:1. This value is greater than the one found by \cite{2011A&A...526A...6W} but lower than the value determined by \cite{1991IAUS..145..375M}.
    
    \item[$-$] According to the distribution in galactic latitude we find evidence that the CSPNe H-rich and H-poor come from different populations, suggesting that H-deficient stars are preferentially formed in younger populations by higher mass progenitors.

\item Our larger sample confirms the existence of a deficiency of [WR]-CSPNe with [WC 5-7] spectral types found by previous authors \cite{2009PhDT.......166T}.
     
     \item[$-$] Upper 80\% of binary CSPNe belong to the H-rich group. We find a significant dearth of H-deficient stars in binary systems. This result suggests that evolutionary scenarios leading to the formation of binary central stars interfere with the conditions required for the formation of H-deficient CSPNe. This is in agreement with the expectations from both merger and late thermal pulse scenarios.
     
     \item We find that 50\% of our sample with derived values of $\log L$, $\log T_{\rm eff}$, and $\log g$ (88 out of 175 objects) shows masses and ages consistent with single stellar evolution models. This suggests that some single stars are indeed able to form PNe.
\end{itemize}

The task of building large catalogues is hard and time-consuming, 
but it is the only way to organize the knowledge achieved by the astronomical 
community and to promote new studies that will make progress in any area of research. 
We hope that the work presented here will constitute a mobilizer of ideas 
and a driving force for new projects that will result in an increase in our 
understanding of the final stages of ordinary mass stars.

%%%%%%%%%%%%%%%%%%%%%%%%%%%%%%%%%%%%%%%%%%%%%%%%%%%%%%%%%%%%%%%%%%%%%%%%%
%%%%%%%%%%%%%%%%%%%%%%%%%%%%%%%%%%%%%%%%%%%%%%%%%%%%%%%%%%%%%%%%%%%%%%%%%
%%%%%%%%%%%%%%%%%%%%%%%%%%%%%%%%%%%%%%%%%%%%%%%%%%%%%%%%%%%%%%%%%%%%%%%%%

\section{Acronyms and bibliographical references of table~\ref{tabla1} and \ref{tabla2}}

AB2015 \cite{2015MNRAS.452.2911A}

AC2007 \cite{2007A&A...472..497A}

AC2016 \cite{2016MNRAS.457.3409A}

AD2016 \cite{2016MNRAS.462.1393A}

AD2019 \cite{2019MNRAS.484.3251A}

AE1985 \cite{1985ApL....24..205A}

AE2008 \cite{2008AstL...34..839A}

AG1985 \cite{1985Ap&SS.113...59A}

AH2020 \cite{2020MNRAS.tmp..425A}

AI2013 \cite{2013AstL...39..201A}

AK1985 \cite{1985PASP...97.1142A}

AK1987 \cite{1987ApJS...65..405A}

AK1988 \cite{1988PASP..100..192A}

Al1975 \cite{1975MSRSL...9..271A}

AL2019 \cite{2019MNRAS.490.2475A}

AL2019a \cite{2019arXiv191109991A}

AM2013 \cite{2013A&A...552A..25A}

AM2014 \cite{2014MNRAS.444.3459A}

AM2015 \cite{2015MNRAS.446..317A}

AM2015b \cite{2015MNRAS.448.2822A}

AN1999 \cite{ 1999AstL...25..615A}

AN2003 \cite{2003A&A...403..659A}

AT2003 \cite{2003ApJS..147...97A}

AW2019 \cite{2019ApJ...882..171A}

B1989 \cite{1989AJ.....97..476B}

B2008 \cite{2008ApJ...674..954B}

BA2016 \cite{2016MNRAS.458.2694B}

bc-CSPN binary for the cool CSPN

BC1999 \cite{1999PASP..111..217B}

BC2003 \cite{2003A&A...400..161B}

BC2012 \cite{2012ApJS..200....3B}

BC2016 \cite{2016ApJ...826..139B}

BD1982 \cite{1982PVSS...10....1B}

BH1994 \cite{1994MNRAS.271..175B}

BJ2018 \cite{2018A&A...619A..84B}

BJ2019 \cite{2019MNRAS.482.4951B}

BK2018 \cite{2018A&A...620A..84B}

BM1986 \cite{1986A&A...168..248B}

BM2000 \cite{2000A&AS..146..407B}

BM2012 \cite{2012Sci...338..773B}

BM2012a \cite{2012A&A...545A.146B}

BO2002 \cite{2002ASPC..279..239B}

BP2002 \cite{2002Ap&SS.279...31B}

BP2003 \cite{2003AJ....125..260B}

C1973 \cite{1973PASP...85..401C}

C1980 \cite{1980AJ.....85..853C}

C1984 \cite{1984ApJ...279..304C}

C1995 \cite{1995MNRAS.276..521C}

CA1996 \cite{1996A&A...307..215C}

CB1999 \cite{1999AJ....118..488C}

CB2016 \cite{2016MNRAS.457....9C}

Cd1993 \cite{1993A&A...276..184C}

Cd1998 \cite{1998MNRAS.296..367C}

Cd2014 \cite{2014AJ....147..142C}

CG2009 \cite{2009AJ....138..691C}

cH1989 \cite{1989IAUS..131..443C}

CJ1987 \cite{1987ApJ...321L.151C}

CK2006 \cite{2006ApJ...646L..69C}

CM2000 \cite{2000RMxAA..36....3C}

CP1985 \cite{1985ApJ...291..237C}

CR2014 \cite{2014MNRAS.441.2799C}

CS2011 \cite{2011MNRAS.410.1349C}

CZ2001 \cite{2001A&A...368..160C}

D1983 \cite{1983ApJ...270L..13D}

D1985 \cite{1985ApJ...294L.107D}

D1999 \cite{1999RvMA...12..255D}

d2006 \cite{2006IAUS..234..111D}

d2009 \cite{2009PASP..121..316D}

D2014 \cite{2014PhDT........76D}

DB1996 \cite{1996ApJ...468L.111D}

dB2004 \cite{2004ApJ...602L..93D}

dC2001 \cite{2001MNRAS.328..527D}

DH1998 \cite{1998A&A...334..618D}

DL1985 \cite{1985ApJ...290..321D}

dM2002 \cite{2002AJ....124..464D}

dO2015 \cite{2015MNRAS.448.3587D}

DP2011 \cite{2011MNRAS.414.2812D}

DP2013 \cite{2013MNRAS.434.1513D}

dP2013 \cite{2013MNRAS.428.2118D}

DW1996 \cite{1996A&A...314..217D}

EP2003 \cite{2003MNRAS.341.1349E}

EP2005 \cite{2005MNRAS.359..315E}

F1994 \cite{1994PASP..106...56F}

F1996 \cite{1996PASP..108..664F}

FB2014 \cite{2014MNRAS.440.1345F}

Fd2008 \cite{2008PhDT.......109F}

FK1983 \cite{1983ApJ...269..592F}

FK2001 \cite{2001ApJS..136..735F}

FL1987 \cite{1987ApJ...316..399F}

FL1999 \cite{1999ApJ...518..866F}

FM1981 \cite{1981ApJ...251..205F}

FM2010 \cite{2010PASA...27..203F}

FP2006 \cite{2006MNRAS.372.1081F}

FP2016 \cite{2016MNRAS.455.1459F}

G2014 \cite{2014A&A...570A..26G}

GB1983 \cite{1983ApJ...271..259G}

GB2010 \cite{2010ApJ...720..581G}

GC2009 \cite{2009A&A...500.1089G}

Gd2013 \cite{2013A&A...553A.126G}

GD2014 \cite{2014AJ....148...57G}

GG2006 \cite{2006RMxAA..42..127G}

GG2014 \cite{2014A&A...567A..12G}

GH2008 \cite{2008ASPC..391..107G}

GJ1989 \cite{1989ApJ...338..862G}

GL1984 \cite{1984ApJ...280..177G}

GM2013 \cite{2013A&A...551A..53G}

GM2019 \cite{2019ApJ...885...84G}

GM2019b \cite{2019A&A...630A.150G}

GP1988 \cite{1988A&A...197..266G}

GP2001 \cite{2001A&A...373..572G}

GP2003 \cite{2003A&A...407.1007G}

GS1997 \cite{1997A&A...318..256G}

GS2003 \cite{2003IAUS..209...43G}

GS2004 \cite{2004A&A...427..231G}

GS2011 \cite{2011A&A...525A..58G}

GS2016 \cite{2016NewA...45....7G}

GT2000 \cite{2000A&A...362.1008G}

GZ2006 \cite{2006A&A...451..925G}

GZ2007 \cite{2007A&A...467L..29G}

GZ2014 \cite{2014A&A...566A..48G}

H1996 \cite{1996ASPC...96..127H}

H1977 \cite{1977ApJ...215..609H}

H1999 \cite{1999PASP..111.1144H}

H2003 \cite{2003IAUS..209..237H}

HA1987 \cite{1987ApJ...313..268H}

HA1994 \cite{1994ApJS...93..465H}

HA1997 \cite{1997ApJ...491..242H}

HA1999 \cite{1999ApJ...514..878H}

HA1999b \cite{1999ApJ...525..294H}

HA2001 \cite{2001AJ....122..954H}

HA2001b \cite{2001PASP..113.1559H}

HB2004 \cite{2004PASP..116..391H}

HB2006 \cite{2006IAUS..234..421H}

HB2010 \cite{2010AJ....140..319H}

HB2011 \cite{2011MNRAS.417.2440H}

HB2015 \cite{2015ApJ...813..121H}

HB2016 \cite{2016AJ....152...34H}

HD2007 \cite{2007AJ....133..631H}

HD2013 \cite{2013ApJ...770...21H}

HF2000 \cite{2000A&A...355L..27H}

HF2015 \cite{2015AJ....150...30H}

HF2017a \cite{2017AAS...22914810H}

HF2017 \cite{2017AJ....153...24H}
 
HH1994 \cite{1994AAS...185.4710H}

HI2004 \cite{2004INGN....8....6H}

HI2006 \cite{2006AJ....131.3040H}

HJ2016 \cite{2016ApJ...832..125H}

HK2008 \cite{2008ApJ...680.1162H}

HK2010 \cite{2010ApJ...725..173H}

HM1990 \cite{1990Ap&SS.169..183H}

HM2008 \cite{2008BaltA..17..293H}

HM2013 \cite{2013AAS...22124909H}

HP1993 \cite{1993ApJ...411L.103H}

HP2007 \cite{2007A&A...467.1253H}

HP2013 \cite{2013MNRAS.434.1505H}

HS2018 \cite{2018MNRAS.473..241H}

Hv2015 \cite{2015A&A...573A..65H}

HW1988 \cite{1988A&A...194..223H}

HZ2010 \cite{2010MNRAS.406..626H}

HZ2019 \cite{2019Ap&SS.364...32H}

JB2014 \cite{2014A&A...562A..89J}

JB2015 \cite{2015A&A...580A..19J}

JB2019 \cite{2019MNRAS.482L..75J}

JE1969 \cite{1969Obs....89...18J}

JH2019 \cite{2019MNRAS.487.4128J}

JP2019 \cite{2019MNRAS.489.2195J}
 
JT1996 \cite{1996A&A...307..200J}

JV2017 \cite{2017A&A...600L...9J}

K1994 \cite{1994AstL...20..644K}

K2012 \cite{2012AstL...38..707K}

KB1994 \cite{1994MNRAS.271..257K}

KB2003 \cite{2003ApJ...591L..37K}

KB2005 \cite{2005A&A...441..289K}

KB2014 \cite{2014MNRAS.442.1379K}

KH1997 \cite{1997A&A...320...91K}

KH1997b \cite{1997IAUS..180..114K}

KJ1991 \cite{1991ApJ...372..215K}

KK2001 \cite{2001ApJ...559..419K}

KK2013 \cite{2013ApJS..204....5K}

KM1997 \cite{ 1997IAUS..180...64K}

Ko2001 \cite{2001Ap&SS.275...41K}

KO2015 \cite{2015MNRAS.453.1879K}

KP2016 \cite{2016MNRAS.455.3413K}

KS1988 \cite{1988Obs...108...88K}

KS1990 \cite{1990ApJ...359..392K}

KS1993 \cite{1993A&A...279..529K}

KV2015 \cite{2015MNRAS.450.3514K}

L1977 \cite{1977A&A....60...93L}

L1984 \cite{1984ApJ...279..714L}

LB2013 \cite{2013ApJ...769...32L}

LC2011 \cite{2011A&A...527A.105L}

LC2014 \cite{2014A&A...563A..43L}

LF1988 \cite{1988PASP..100..187L}

LF2010 \cite{2010PASP..122..524L}

LF2015 \cite{2015A&A...579A..39L}

LH1994 \cite{1994A&A...283..567L}

LH1998 \cite{1998A&A...330..265L}

LH2005 \cite{2005A&A...430..223L}

LH2013 \cite{2013A&A...549A..65L}

LK1987 \cite{1987BAAS...19.1090L}

LR1983 \cite{1983A&A...123...33L}

LR2019 \cite{2019MNRAS.489.1054L}

LS2000 \cite{2000MNRAS.312..585L}

LS2007 \cite{2007AJ....133..987L}

LT1995 \cite{1995ApJ...441..424L}

LZ1998 \cite{1998A&A...340..117L}

M1989 \cite{1989IAUS..131..261M}

M1990 \cite{1990A&A...229..152M}

M1991 \cite{1991IAUS..145..375M}

M1997 \cite{1997ARep...41..760M}

M2012 \cite{2012MNRAS.423..934M}

MA2003 \cite{2003AJ....126..887M}

MA2009 \cite{2009A&A...496..813M}

MASH-I \cite{2006MNRAS.373...79P}

MASH-II \cite{2008MNRAS.384..525M}

MB2012 \cite{2012MNRAS.419...39M}

MB2013 \cite{2013MNRAS.428L..39M}

MB2013b \cite{2013MNRAS.436.3068M}

MC2006 \cite{2006A&A...458..203M}

MC2011 \cite{2011MNRAS.413.1264M}

MG1988 \cite{1988A&A...197L..25M}

MH1991 \cite{1991A&A...252..265M}

MJ2011 \cite{2011A&A...531A.158M}

MK1985 \cite{1985A&A...142..289M}

MK1988 \cite{1988A&A...190..113M}

MK1992 \cite{1992A&A...260..329M}

MK2015 \cite{2015ApJ...800....8M}

MK2016 \cite{2016ApJ...829...73M}

ML2015 \cite{2015MNRAS.451..870M}

MM2011 \cite{2011A&A...528A..39M}

MM2013 \cite{2013MNRAS.432.3186M}

MM2014 \cite{2014MNRAS.440.1410M}

MM2015 \cite{2015MNRAS.448.1789M}

MM2018a \cite{2018MNRAS.473.2275M}

MM2018 \cite{2018PASA...35...27M}

MM2019 \cite{2019PASA...36...18M}

MO2007 \cite{2007IAUS..240..429M}

MP2001 \cite{2001MNRAS.322..877M}

MP2006 \cite{2006RMxAA..42...53M}

MP2007 \cite{2007MNRAS.374.1404M}

MP2013 \cite{2003MNRAS.346..719M}

MP2013b \cite{2013MNRAS.435..606M}

MR1991 \cite{1991ApJ...371..380M}

MS2012 \cite{2012BaltA..21..180M}

MT2001 \cite{2001MNRAS.321..487M}

MV1997 \cite{1997MNRAS.288..777M}

MV2010 \cite{2010PASA...27..199M}

MV2016 \cite{2016A&A...593A..29M}

Mv2019 \cite{2019MNRAS.487.1040M}

MW1990 \cite{1990MNRAS.247..177M}

MW1993 \cite{1993A&A...268..561M}

MW2016 \cite{2016MNRAS.456..633M}

MZ2002 \cite{2002A&A...383..188M}

N1999 \cite{1999A&A...350..101N}

NA2000 \cite{2000A&A...358..321N}

NH1994 \cite{1994A&A...292..239N}

NS1993 \cite{1993IAUS..155..495N}

NS1995 \cite{1995A&A...301..545N}

NT2005 \cite{2005AIPC..804..173N}

OH2015 \cite{2015ApJS..217...22O}

OH2019 \cite{2019arXiv191101390O}

OK2013 \cite{2013ApJ...764...77O}

OK2014 \cite{2014MNRAS.437.2577O}

OK2014b \cite{2014A&A...565A..87O}

P1983 \cite{1983IAUS..103..323P}

P1984 \cite{1984ASSL..107.....P}

P1996 \cite{1996A&A...307..561P}

P1999 \cite{1999MNRAS.304..127P}

P2003 \cite{2003MNRAS.344..501P}

P2004 \cite{2004A&A...413.1009P}

P2005 \cite{2005RMxAA..41..423P}

PA1989 \cite{1989A&AS...81..309P}

PA1991 \cite{1991A&AS...88..121P}

PA1998 \cite{1998A&A...329L...9P}

PB1994 \cite{1994MNRAS.267..452P}

PB2005 \cite{2005A&A...436..953P}

PB2007 \cite{2007A&A...471..865P}

PB2010 \cite{2010A&A...509A..13P}

PC2005 \cite{2005MNRAS.357..548P}

PC2010 \cite{2010A&A...521A..26P}

PF2004 \cite{2004PASA...21..334P}

PG1995 \cite{1995A&A...300L..25P}

Ph2005 \cite{2005MNRAS.362..847P}

Pi2005 \cite{2005MNRAS.357..619P}

PL2016 \cite{2016AJ....151...53P}

PM2002 \cite{2002RMxAA..38...23P}

PM2003 \cite{2003MNRAS.341..961P}

PM2003b \cite{2003RMxAC..18...84P}

PM2008 \cite{2008A&A...477..535P}

PR1989 \cite{1989RMxAA..17...25P}

PR1997 \cite{1997A&A...317..911P}

PR2013 \cite{2013RMxAA..49...87P}

PR2013b \cite{2013ApJ...771..114P}

PS1998 \cite{1998A&A...337..866P}

PS2005 \cite{2005A&A...429..993P}

PT1992 \cite{1992A&A...265..757P}

R1987 \cite{1987ApJ...315..234R}

RC2001 \cite{2001A&A...377.1042R}

RD1998 \cite{1998A&A...338..651R}

RF2018 \cite{2018Galax...6...88R}

RH2002 \cite{2002A&A...381.1007R}

RK1994 \cite{1994A&A...286..543R}

RK1996 \cite{1996A&A...310..613R}

RK1999 \cite{1999A&A...347..169R}

RP1997 \cite{1997A&AS..126..297R}

RR2014 \cite{2014A&A...565A..40R}

RR2017 \cite{2017MNRAS.464L..51R}

S2002 \cite{2002A&A...390..667S}

SB2007 \cite{2007AJ....134..846S}

SB2008 \cite{2008ARep...52..558S}

SB2019 \cite{2019A&A...630A..80S}

SB2019b \cite{2019arXiv191009680S}

SC1993 \cite{1993A&A...276..463S}

SECGPN \cite{1992secg.book.....A}

SF1987 \cite{1987A&AS...67..541S}

SG2003 \cite{2003A&A...406..305S}

SH2014 \cite{2014A&A...570A..88S}

SK1989 \cite{1989ApJS...69..495S}

SK1994 \cite{1994A&A...291..604S}

SK1995 \cite{1995A&A...302..211S}

SL2004 \cite{2004A&A...422..563S}

SP1995 \cite{1995ApJ...452..286S}

SP2015 \cite{2015ApJS..218...25S}

SR2015 \cite{2015Natur.519...63S}

SV2002 \cite{2002ApJ...576..285S}

SW1997 \cite{1997A&A...328..598S}

SW2009 \cite{2009ApJ...707L..32S}

SZ1997 \cite{1997MNRAS.289..665S}

TA1989 \cite{1989A&AS...77...39T}

TA1991 \cite{1991A&AS...89...77T}

TA1993 \cite{1993A&AS..102..595T}

TA2010 \cite{2010RMxAA..46..221T}

TE2002 \cite{2002MNRAS.334..875T}

TG2015 \cite{2015ApJ...799...67T}

TH2005 \cite{2005ASPC..334..325T}

TJ2013 \cite{2013MNRAS.436.2082T}

TN2004 \cite{2004ApJ...616..485T}

TP2010 \cite{2010A&A...515A..83T}

TR2019 \cite{2019MNRAS.485.3360T}

TS1987 \cite{1987PASP...99.1264T}

TS1994 \cite{1994A&A...288..897T}

TS2012 \cite{2012AJ....144...81T}

TY2010 \cite{2010ApJ...714..178T}

UL2014 \cite{2014A&A...565A..36U}

vH1981 \cite{1981SSRv...28..307V}

vH1981 \cite{1981SSRv...28..227V}

VH2014 \cite{2014A&A...563L..10V}

W1992 \cite{1992LNP...401..273W}

W1995 \cite{1995BaltA...4..340W}

W2009 \cite{2009PhDT.......208W}

WB2011 \cite{2011MNRAS.418..370W}

WD1995 \cite{1995A&A...298..567W}

WG1985 \cite{1985ApJS...58..379W}

WG2018 \cite{2018A&A...614A.135W}

WG2011a \cite{2011A&A...526A...6W}

WG2011b \cite{2011A&A...531A.172W}

WH2006 \cite{2006PASP..118..183W}

WJ2018 \cite{2018MNRAS.480.4589W}

WK1992 \cite{1992LNP...401..288W}

WK1997 \cite{1997A&A...323..963W}

WL2007 \cite{2007MNRAS.381..669W}

WM2015 \cite{2015A&A...579A..86W}

WO1994 \cite{1994AcA....44..407W}

WR1994 \cite{1994A&A...284L...5W}

WR2016 \cite{2016A&A...593A.104W}

WR2019 \cite{2019MNRAS.483.5291W}

WW1993 \cite{1993A&A...275..256W}

WW1996 \cite{1996A&A...315..253W}

WW2010 \cite{2010A&A...524A...9W}

YL2013 \cite{2013MNRAS.436..718Y}

ZA1986 \cite{1986ApJ...301..772Z}

ZF2012 \cite{2012yCat.1322....0Z}

ZK1993 \cite{1993ApJS...88..137Z}

ZL2002 \cite{2002MNRAS.337..499Z}

ZP1990 \cite{1990A&AS...82..273Z}

ZR2012 \cite{2012IAUS..283..211Z}

ZRW2012 \cite{2012A&A...548A.109Z}

%%%%%%%%%%%%%%%%%%%%%%%%%%%%%%%%%%%%%%%%%%%%%%%%%%%%%%%%%%%%%%%%%%%%%%%%%
%%%%%%%%%%%%%%%%%%%%%%%%%%%%%%%%%%%%%%%%%%%%%%%%%%%%%%%%%%%%%%%%%%%%%%%%%
%%%%%%%%%%%%%%%%%%%%%%%%%%%%%%%%%%%%%%%%%%%%%%%%%%%%%%%%%%%%%%%%%%%%%%%%%

\begin{acknowledgements}

We would like to thank our anonymous referee whose
valuable comments helped us to improve this paper.
Part of this research was supported 
by grant SeCyT UNC project NRO PIP: 33820180100080CB and 
by grant CONICET project NRO: PICT 2017-3301.
W.W. would like to thank Roberto Méndez.
This research has made use of NASA's Astrophysics Data System Service.
This research has made use of the SIMBAD database, operated at CDS, 
Strasbourg, France 2000,A\&AS,143,9, "The SIMBAD astronomical database"
Wenger et al.
This research has made use of the VizieR catalogue access tool, CDS, 
Strasbourg, France (DOI: 10.26093/cds/vizier). The original description of the 
VizieR service was published in A\&AS 143, 23.
\end{acknowledgements}

%%%%%%%%%%%%%%%%%%%%%%%%%%%%%%%%%%%%%%%%%%%%%%%%%%%%%%%%%%%%%%%%%
%%%%%%%%%%%%%%%%%%%%%%%%%%%%%%%%%%%%%%%%%%%%%%%%%%%%%%%%%%%%%%%%%
%%%%%%%%%%%%%%%%%%%%%%%%%%%%%%%%%%%%%%%%%%%%%%%%%%%%%%%%%%%%%%%%%
%%%%%%%%%%%%%%%%%%%%%%%%%%%%%%%%%%%%%%%%%%%%%%%%%%%%%%%%%%%%%%%%%

%\begin{thebibliography}{}
%\end{thebibliography}

\bibliographystyle{aa}
\bibliography{aa3}
%%%%%%%%%%%%%%%%%%%%%%%%%%%%%%%%%%%%%%%%%%%%%%%%%%%%%%%%%%%%%%%%%
%%%%%%%%%%%%%%%%%%%%%%%%%%%%%%%%%%%%%%%%%%%%%%%%%%%%%%%%%%%%%%%%%
%%%%%%%%%%%%%%%%%%%%%%%%%%%%%%%%%%%%%%%%%%%%%%%%%%%%%%%%%%%%%%%%%
%%%%%%%%%%%%%%%%%%%%%%%%%%%%%%%%%%%%%%%%%%%%%%%%%%%%%%%%%%%%%%%%%

\begin{appendix}

\section{Symbiotic stars?}
\label{sec-syst}

Symbiotic stars (SySt) are binary systems composed by a cold giant (red giant or Mira star) and an evolved hot star (in most cases, a white dwarf), in which the giant transfers material to the white dwarf star via stellar wind \citep{2010arXiv1011.5657M}. The wind is ionized by the UV radiation field from the evolved companion, producing a spectrum with emission-lines typical of PNe. In addition to these emission-lines, SySt generally exhibit absorption features (e.g. TiO and VO) produced in the cool stellar photosphere of the giant \citep{Corradi:2003}. Symbiotic stars allow to study physical processes such as the powering mechanism of supersoft X-ray sources \citep{Jordan:1996}, the thermonuclear outbursts \citep{Munari:1997}, the collimation of stellar winds and formation of jets \citep{Tomov:2003} or their relation with the formation of bipolar PNe \citep{Corradi:2003}.

Another important aspect is that SySt are, probably, progenitors of Type~Ia supernovae (see e.g. \citealp{Whelan:1973, Hachisu:1999, Lu:2009, Stefano:2010, Chen:2011}).
SySt are divided into three categories: (\textit{i}) The S-type (\textit{stellar}) in which the giant has a M-spectral type and dominates the emission in the near-IR, showing the presence of stellar photospheres at (3\,000-4\,000 K) 
\citep{Belczy:2000, Akras:2019a}. (\textit{ii}) The D-type systems (\textit{dusty}) in which the cool companion is a Mira variable star and the near-IR emission corresponds to the dusty envelope around the systems. And (\textit{iii}) D'-type symbiotic stars which are characterized by a F, G or K type cool giant surrounded by a dust shell \citep{Allen:1984}. A new type of SySt namely S+IR, a S-type with an infrared excess in the 
11.6 and/or 22.1 $\mu$m bands, was proposed by \citet{Akras:2019a}.

Evolutionary links between PNe and symbiotic systems are likely. A kind of symbiotic outflow around D-type symbiotic Miras shows morphological similarities with bipolar PNe, indicating that symbiotic Miras, in an evolutionary phase, are a prelude to formation of genuine PNe. This transition occurs when the Mira variable of SySt loses its envelope and forms a PN  \citep{Corradi:2003}. Note that nebulae observed around symbiotic Miras are not PNe. On the other hand, many current symbiotic stars may have gone through the PN phase, when the current WD companion was at an early phase in its evolution. Similarly, many PNe with binary central stars may turn-up into symbiotic systems in the future when the companion star evolves into to AGB phase \citep{Frew:2010}.

Symbiotic stars can be discriminated from PNe using either near-IR colours or, in the optical spectra, the presence of the red continuum and strong [\ion{O}{iii}] $\lambda$4363 relative to $\lambda$5007 in higher-excitation objects. Given that symbiotic stars are generally much denser than even the youngest PNe, \citet{Gutierrez:1988} and \citet{Gutierrez:1995} used the [\ion{O}{iii}] $\lambda$4363 / $\lambda$5007 line ratio to separate PNe from symbiotics. Colour-colour diagrams using near-IR and optical colours were used \citep{Schmeja:2001, Ramos-Larios:2005, Corradi:2008}. Recently, \cite{2017A&A...606A.110I} also proposed a number of new diagnostic diagrams in the optical regime for discriminating SySt from PNe. For  instance, they used diagnostic diagrams based on the \ion{He}{i} recombination lines. 

Our catalogue has eleven CSPNe classified as SySt, which are presented in Table~\ref{syst}. Two objects of this list are classified as possible PNe (PM~1$-$322 and Mz~3) in the Hong Kong/AAO/Strasbourg H$\alpha$ %\ha{} 
PNe database (HASH; \citealp{Parker:2016}). The spectroscopy of these possible PNe is insufficiently conclusive due to the combination of low S/N spectra, low surface brightness and insufficient wavelength coverage. The spectrum of PM~1$-$322 presents the typical forbidden emission lines of a PN as well as the Balmer and Paschen recombination lines. The spectral characteristics indicate that PM~1$-$322 is a young PN, however, its high density and its position on the diagnostic diagram of \citet{Gutierrez:1995} suggest a symbiotic nature \citep{Pereira:2005}. Mz 3 is widely classified as a bipolar PN based on its optical spectral characteristics and morphology \citep{Acker:1992}. However, the position of Mz 3 on the near-IR colour diagrams strongly indicates that it has a symbiotic binary nucleus \citep{Schmeja:2001}.

M~1$-$44 and CTSS~2 are catalogued as true PNe in the HASH catalogue. They are confirmed PNe with multi-wavelength PN-type morphologies and PN spectral features \citep{Parker:2016}. For M~1$-$44 the near-IR photometry indicates the possible presence of a cool component (G-K giant). However, the giant may not be associated with the nebula \citep{2000A&AS..146..407B, Phillips:2007}. 

The source K~4$-$57 is presented as a PN candidate in HASH \citep{Parker:2016}. It appears in the PNe catalogue by \citet{Acker:1992}. However, its optical spectrum is similar to a SySt with nebular emission lines of low to high ionization species and a faint slow rising continuum \citep{Rodriguez:2014}. 
In addition, the location of the source in the  high-density region of the optical diagram from \cite{Gutierrez:1995}, which is typical of symbiotic stars, reveals the possible presence of a symbiotic nebula.

Hen~2$-$442 was previously classified as PN \citep{Acker:1992}, however it is now catalogued as a true SySt \citep{Parker:2016}. The near-IR spectrum exhibits the presence of a red giant and the optical one is dominated by nebular lines and shows a weak evidence  of TiO bands \citep{Yudin:1983}, confirming its symbiotic nature \citep{Rodriguez:2014}.

The other five objects: H~2$-$43, Th~4$-$1, Hen~2$-$25, Hen~2$-$57 and 19w32 are classified as SySt candidates \citep{Parker:2016}. Three of these objects (H~2$-$43, He~2$-$25, and 19w32) were previously classified as planetary nebulae. They were subsequently found to have near-IR colours which are consistent with symbiotic binary nucleus \citep{2000A&AS..146..407B, Schmeja:2001, Pereira:2005, Phillips:2007}. The spectra of Th~4$-$1 and Hen~2$-$57 indicate they are genuine PNe, but this spectroscopic information do not allow to totally rule out the possibility of  being a symbiotic nature.

Table~\ref{syst} shows that these objects, in general, belong to the symbiotic class D/D' \citep{Akras:2019a}. D-type SySt are in more evolved phase than S-types. D-types contain AGB stars as the cold components whereas the S-type SySt have RGB stars \citep{Muerset:1996}. The intense mass-loss generated during the end of the TP-AGB stage is responsible for the formation of the dusty shell in D-type SySt.

\begin{table}[ht]
\caption{Sample of CSPNe classified as SySt.} % title of Table
\centering % used for centering table
\begin{tabular}{c c c c} % centered columns (4 columns)
\hline %inserts double horizontal lines
PN~G & name & prev. class.  & Akras class. \\ [0.5ex] % inserts table
%heading
\hline % inserts single horizontal line
\#8          & PM~1$-$322 & ? & \hspace{0.1cm}D'  \\
003.4$-$04.8 & H~2$-$43   & D & D  \\
004.9$-$04.9 & M~1$-$44   & S & S$+$IR  \\
007.5$+$04.3 & Th~4$-$1   & ? & ?  \\
044.1$+$05.8 & CTSS~2     & ? & ?  \\
061.8$+$02.1 & Hen~2$-$442 & D & D  \\
107.4$-$00.6 & K~4$-$57   & D & D  \\
275.2$-$03.7 & Hen~2$-$25  & ? & \hspace{0.1cm}D'  \\
289.6$-$01.6 & Hen~2$-$57  & ? & D  \\
331.7$-$01.0 & Mz~3       & ? & D  \\
359.2$+$01.2 & 19w32      & D & \hspace{0.1cm}D' \\
\hline \hline 
\end{tabular}
\label{syst} 
\end{table}

%%%%%%%%%%%%%%%%%%%%%%%%%%%%%%%%%%%%%%%%%%%%%%%%5
%%%%%%%%%%%%%%%%%%%%%%%%%%%%%%%%%%%%%%%%%%%%%%%%ç
%%%%%%%%%%%%%%%%%%%%%%%%%%%%%%%%%%%%%%%%%%%%%%%%

\section{Objects rejected from the catalogue}

IRAS~17150$-$3224 (PN~G353.8$+$02.9),
strong evidence that is not PN.
\cite{1993A&A...273..185H} show the optical spectra
and classify it as G2 I.

RE~1738$+$665 (PN~G096.9$+$32.0), it is a
faint, irregular HII region in the ISM
\citep{2008PhDT.......109F}.

\cite{1996A&A...312L..21Z}  proved that the PNe
		He~2$-$436 (PN~G004.8$-$22.7) and 
		WRAY~16$-$423 (PN~G006.8$-$19.8)
 are not linked to our galaxy.

Sh~2$-$128 (PN~G325.8$+$04.5),
\cite{2003AJ....126.1861B} showed that this is an HII region.

%%%%%%%%%%%%%%%%%%%%%%%%%%%%%%%%%%%%%%%%%%%%%%%%%%%%%%%%%%%%%%%%%%%%%%%%%
%%%%%%%%%%%%%%%%%%%%%%%%%%%%%%%%%%%%%%%%%%%%%%%%%%%%%%%%%%%%%%%%%%%%%%%%%
%%%%%%%%%%%%%%%%%%%%%%%%%%%%%%%%%%%%%%%%%%%%%%%%%%%%%%%%%%%%%%%%%%%%%%%%%

\section{Spectral classification of peculiar CSPNe}\label{peculiarCSPN}

Thanks to the identification of certain ions in published spectra, 
it was possible to improve the spectral classification for several objects. 
In the cases when two or more different
spectral classifications were available, the relevance of each one is discussed in the text.

2MASS~J18482874$-$1237434: 
This object was classified as PN by \cite{2014MNRAS.440.1410M}
who reported 
\ion{He}{ii} absorption lines at 4540 and 4686\AA. According to our spectral 
classification system, this CSPN is an O-type.

A~41:
This object requires  special attention. On one hand,
\cite{1994AcA....44..407W}
reported that the primary component of the binary system
is a sdB (they do not show spectra). On the other hand,
\cite{1984ApJ...280..177G}
classified this CSPN as sdO, and show spectra
where H$\beta$ is clearly visible in absorption.
As a more probable classification, we prefer a sdO type.

A~46:
According to \cite{1994MNRAS.267..452P}, this CSPN presents Balmer, \ion{He}{ii}, and 
\ion{N}{v}
$\lambda$4603$-$19 lines in absorption. It also shows emission of
\ion{C}{iv}, %at 5806
\ion{C}{iii}, and
\ion{N}{iii}.
On the other hand, \cite{2007MNRAS.374.1404M} classified it as M6 V.
Moreover, \cite{2009PASP..121..316D} report that the secondary
star is a M6 V.
There exists a possibility that the object is a sdO
\citep{2015MNRAS.446..317A}.
We classify this CSPN as O(H)3 + M6 V.

CRBB~1:
According to the spectra shown by \cite{1991ApJ...371..380M}
we classify this star as B0 III.

ESO~330$-$9:
This object presents a distorted morphology. The CSPN is not at the geometric center.
\cite{2017AJ....153...24H} show a spectrum
where only an absorption of \ion{He}{ii} at 4686\AA\ is visible.
Moreover, Fig.~1 of 
\cite{2012Sci...338..773B} shows clear Hydrogen absorption lines, \ion{He}{ii} and
\ion{N}{v} (4603-19\AA) together with
emission lines of \ion{C}{iv} and \ion{O}{v}\footnote{\url{https://science.sciencemag.org/content/sci/suppl/2012/11/07/338.6108.773.DC1/Boffin.SM.pdf}}.
In addition, the authors report that this planetary nebula nucleus 
is constituted by a double-degenerate central star. 
Both stars are responsible for the ionization of the nebula.
We classify this CSPN as O(H)3-4.

H~2$-$1:
The CSPN is classified as [WC 11] by
\cite{2009A&A...500.1089G},
who show a spectrum in the range
5200-7300\AA.
However, in the spectrum taken by
 \cite{1988A&A...190..113M} H$\gamma$ is in absorption, 
 \ion{He}{i}, and \ion{He}{ii} are evident.
We reclassify this CSPN as O(H)5-9.

Hb~12:
\cite{1985PASP...97.1142A}
suggest that the emission lines equivalent width does not correspond to a [WR].
The H$\alpha$ line presents broad emission \citep{2002RMxAC..12..154A}
and the spectrum does not present absorption lines \citep{2003ApJS..147...97A}.
This object could be in a previous stage to a PN.
We keep both classifications,
possible B[e] and possible [WN 7].

Hen~2$-$11:
\cite{2003IAUS..209...43G} reported non-stellar emission lines,
but \cite{2004A&A...427..231G} classified this object as wels. 
However, the spectrum showed by \cite{2014A&A...562A..89J}
displays subtle \ion{He}{ii} absorption at 5412\AA.
In this context, we prefer an O-type.

Hen~2$-$108:
\cite{2014A&A...570A..26G} classified this CSPN as VL. 
However, the spectra display clear \ion{He}{ii} absorption at 5412\AA\ 
and subtle \ion{C}{iv} absorption at 5806\AA. 
According to the spectrum shown by \cite{2007A&A...467.1253H}, the CSPN shows 
\ion{He}{i}, \ion{He}{ii}, and H$\gamma$ absorption. 
We classify this object as O(H).

Hen~2$-$131:
\cite{2014A&A...570A..26G} classified this CSPN as VL. 
However, the spectra display clear \ion{He}{ii} absorption  
at 5412\AA\ and subtle  \ion{C}{iv} absorption at 5806\AA. 
\cite{ 1986ApJ...301..772Z} reported a  Of8 spectral type and P-Cygni profile.
\cite{1977ApJ...215..609H} reported a O7(f)eq spectral type and 
showed the spectra where Balmer line absorption is visible.
While \cite{2013A&A...553A.126G} reported Of(H) and
 \cite{1993A&AS..102..595T} classified it as wels.
We prefer to adopt a spectral type O(H).

Hen~2$-$138:
\cite{1988A&A...190..113M}
classified this object as Of(H),
but considering these author spectra we propose a re-classification 
as O(H)7-9f.

Hen~2$-$146:
There are two possible CSPNe in this object. 
Both display a late spectral type 
\citep{1969Obs....89...18J}.
To be consistent with this catalogue nomenclature, 
we prefer a classification K-M.

Hen~2$-$155:
The spectra showed by \cite{2015A&A...580A..19J}
display H absorption lines, \ion{He}{ii}, 
\ion{O}{v} (at 5114\AA), and \ion{C}{iv}  (at 5801-12\AA).
We classify this CSPN as O(H)3-5.

Hen~2$-$161:
The spectra show Balmer absorption lines, \ion{He}{ii} $\lambda$4542\AA, 
\ion{He}{ii} $\lambda$5412\AA, 
\ion{N}{v}   $\lambda$4603-19\AA,
\ion{O}{v} at $\lambda$5114, and \ion{C}{iv}
emission at $\lambda$5801-12\AA\
(\citet{2015A&A...580A..19J} and private communication with Jones). 
In addition I(4542)$\sim$I(4603), in this sense we suggest
a spectral type O(H)3-4.

Hen~2$-$442:
 \cite{1985ApL....24..205A}
reported that there are two stellar sources separated by 6~arcsec,
and both sources produce PN-like emission spectra.
A description of this objects is presented by
 \cite{2014A&A...567A..49R}.

Hen~2$-$260:  
According to the spectrum shown by \cite{2007A&A...467.1253H}, 
the CSPN shows \ion{He}{i}, \ion{He}{ii}, and Balmer absorption. 
I(4471)$>$I(4542) implies a later star than O(H)7. 
Nevertheless, the intensity of the \ion{He}{ii} lines is significant. 
\ion{Si}{iii} absorption lines (at 4552, 4567 and 4574\AA) 
are evident in the spectrum published by \cite{1997IAUS..180...64K}. 
According to this, we classify this star as O(H)7-8.

Hf~2$-$2:
\cite{2016AJ....152...34H}
indicate that Balmer and \ion{He}{ii} absorption lines are clearly visible in
the spectrum.
Moreover, in the spectrum shown by the authors,  strong \ion{N}{v} 
absorption lines at 4603-19\AA\ are present. This features are typical of an O(H)3 star.

IC~2149:
This CSPN is classified as O4f in SECGPN,
and \cite{2013A&A...553A.126G} classify it as
Of(H).
 \cite{1994ApJ...426..653F} reported
stellar features at the optical range.
We adopt the spectral type O(H)4f.

IC~2553:
\cite{2017PASA...34...36A}
showed that the emission lines of
\ion{N}{iii} 4631+4641\AA, 
\ion{C}{iii} 4650\AA, and
\ion{O}{iii}$+$\ion{O}{v}  5592\AA\
are of nebular origin.
The only line that could be stellar
is 5806\AA.
These authors ruled out the wels classification.
In this sense, we prefer to classify this CSPN as
emission line.

IC~4776:
\cite{1975MSRSL...9..271A}
classified this CSPN as a possible [WC 6].
More recently, \cite{2019MNRAS.487.1040M}
 analyzed and presented high resolution spectra.
They ruled out a spectral type [WR] neither Of. 
Moreover, they could not affirm if the CSPN is 
H-rich or not.
According to the strong \ion{He}{ii} absorption line
we prefer an  O-type spectral classification.

IC 4846:
This CSPN displays H$\gamma$ absorption lines and 
\ion{He}{ii} (4200, 4541\AA) 
together with  \ion{He}{ii} emission line at 4686\AA \ \citep{2001PASP..113.1559H}.
In addition \cite{2014A&A...567A..12G} reported  \ion{C}{iv} emission at 5806\AA.
We adopt a classification O(H)3-4 f.

IC~5217:
Certainly the central star of this object requires further study. \cite{1985ApJ...291..237C}
classified this object as Of-WR,
\cite{2001AJ....122..954H} as WC(WNb ?),
\cite{1993A&AS..102..595T} as wels, and
\cite{2003A&A...403..659A} as [WC 8-9]?.
There is no spectrum published.
Cautiously, we adopt a classification [WC]?.

K~1$-$2:
\cite{2003MNRAS.341.1349E}
described this object but does not show the spectrum or classify it.
They detected \ion{He}{ii} and H$\beta$ stellar features.
In addition, they mentioned that H$\beta$ absorption is wide.
On the other hand, \cite{2009PASP..121..316D} reported 
that the secondary star of the binary system is a K2 V (or earlier).
We classify it as O(H)~+~K2~V.

M~1$-$12:
Previously, in \cite{2011A&A...526A...6W}
we classified this CSPN as [WC 10-11].
Nevertheless, in a spectra showed by \cite{2012AstL...38..707K}
an absorption at 5412\AA\ is clear.
So, we prefer a classification O-type.

M~1$-$37:
 According to the spectrum shown by \cite{2007A&A...467.1253H}, the CSPN shows \ion{He}{i},
 \ion{He}{ii}, and H$\gamma$ absorption. 
 Taking into account that it is a noisy spectrum, we can say that the star is O(H) type. 
 The spectrum presented by  \cite{1997IAUS..180...64K} shows strong P-Cygni-type  profiles at 4471\AA. 
 Although this object was classified as [WC 11] by \cite{2006A&A...451..925G}, 
 this classification is discarded because of the H$\gamma$ feature. On the other hand, 
 the object was classified as a peculiar star by HP2007.

M~1$-$38:
\cite{2009A&A...500.1089G} classified this object as [WC 11],
but its spectra is not conclusive.
On the other hand, \cite{2007A&A...467.1253H}
showed an spectrum with clear Balmer, \ion{He}{i}, and \ion{He}{ii}
absorption lines. Moreover, I(4471)$>$I(4542).
We classify this star as  O(H)7-8.

M~1$-$44:
Seemingly the object is not well identified.
\cite{ 1983PASP...95..739L}
did one of the first identifications and mentioned two stars 
that are separated by 2~arcsec.
Later, \cite{ 1987A&AS...68...51S} classified this star as
possible PN,
and possible SySt is the conclusion of \cite{2000A&AS..146..407B}.
A deeper investigation is required for this object.

M~1$-$77:
Object classified as possible OB star by SECGPN.
\cite{ 2009PASP..121..316D} showed some absorption lines, i.e.,
H$\delta$, \ion{O}{ii},  and \ion{He}{i}.
This information is not enough to classify it.
We assume these are the most intense lines in the spectrum.
The \ion{O}{ii} features are especially important in B1 type star
\citep{1990PASP..102..379W}. Hence, we classify this star as B1.

M~2$-$12:	 
According to the spectrum shown by \cite{2007A&A...467.1253H},
the CSPN displays \ion{He}{i} and \ion{He}{ii} absorption lines, with
I(4471)$>$I(4542).
The presence of Balmer features is not evident.
\cite{2004A&A...427..231G} classified this objects as [WC 11], 
although the spectrum shown by \cite{2007A&A...467.1253H} discards this classification.
We classify this star as an O7-8.

M~2$-$54:
It is a variable star with a magnitude variation 
amplitude of up to 0.3 mag in Johnson V band  \citep{1999A&AS..135..493H}.
It could also be a post-AGB star
\citep{2003A&A...407.1007G}.
Nevertheless, the same authors classify it as O9 V.

M~3$-$1:
We could not detect any stellar feature
in this object
\citep{2011A&A...526A...6W}.
But in the spectra showed by
\cite{2019MNRAS.482L..75J} absorptions at H$\beta$ and 5412\AA\ are evident. 
We classify this CSPN as O(H).

M~3$-$2:
Spectra showed by \cite{2012AstL...38..707K}
display clear Balmer absorption lines.
We classify this CSPN as H-rich.

My~60:
\cite{2016MNRAS.458.2694B} classified this CSPN as O(H), however 
they only observed \ion{O}{vi} emission lines at 3811-34\AA. 
On the other hand, \cite{2014A&A...570A..26G} has classified it as wels.
In this situation we prefer to classify it as emission line.

MyCn~18:
This object received different types of classifications,
\cite{1991IAUS..145..375M} and
\cite{2018PASA...35...27M} classified the CSPN as Of(H).
On the other hand,
\cite{2013RMxAA..49...87P} report a spectral type
[WC]$-$PG~1159.
While \cite{2007AJ....133..987L}
with high quality spectra did not detect Balmer
absorption and 
classified this object as Of(c).
Based on the quality of the published spectra we
choose the classification Of(c).

NGC~1360:
CSPN classified as O(H) by
\cite{1991IAUS..145..375M}.
Nevertheless, in the spectra showed by
\cite{1988A&A...190..113M} strong absorption of
\ion{He}{ii} at 4686\AA\ is visible.
On the other hand, in the spectra presented by
\cite{2011MNRAS.417.2440H}
no \ion{He}{i} absorption can be appreciated.
In this sense, we classified this CSPN
as O(H)3-4.

NGC~1535:
In the spectrum showed by
\cite{1988A&A...190..113M}
 the H absorption is evident, and it is
confirmed by \cite{2013A&A...553A.126G}.
According to this, we classify this CSPN as O(H)5.

NGC~3211:
\cite{2016MNRAS.458.2694B} classified this CSPN as O(H). 
Nevertheless, they did not observe the Balmer series; they
only detected  \ion{O}{vi} emission lines at 3811-34\AA.
In this situation we prefer to classify this object as emission line.

NGC~5979:
\cite{2016MNRAS.458.2694B} observed this CSPN and classified it as O (H). 
However, according to the features
displayed in their spectra, in particular \ion{C}{iv}  at 5806\AA\ 
and \ion{O}{iii}  at 5592\AA, we prefer to classify this object as O(H)3-4.
It is noteworthy that the spectrum obtained by \cite{2016MNRAS.458.2694B} 
shows clear  \ion{O}{vi} emission lines. 
These lines are typical of [WO] stars.

NGC~6026:
WD2011a reported a spectral type O(H)7. For their part,
\cite{2010AJ....140..319H} classified this object as O7 V.
So we classify this CSPN as O(H)7 V.

NGC~6058:
Object classified previously by \cite{1984ASSL..107.....P}
as O9, and by \cite{2013A&A...553A.126G} as O(H).
Nevertheless, in the spectra showed by
\cite{2011MNRAS.417.2440H}
strong H, \ion{He}{ii}, and \ion{N}{v} absorptions are revealed.
We classify this CSPN as O(H)3.

NGC~6210:
Object classified  by \cite{1978A&A....62...95P}
as O6, and as O(H) by \cite{2013A&A...553A.126G}.
Nevertheless, in the spectra showed by
\cite{2011MNRAS.417.2440H} strong H, \ion{He}{ii}, and \ion{N}{v} absorption are revealed.
We decided to classify this CSPN as O(H)3.

NGC~6629:
This CSPN was classified as
possible [WC 4]
\citep{2003A&A...403..659A}
and as wels by \citep{1993A&AS..102..595T}.
Nevertheless, in the spectra showed by
\cite{1988A&A...190..113M} and
\cite{2014A&A...563A..43L},
the Balmer series together with the \ion{He}{ii} absorption are present.
According to this, we propose a spectral type O(H) for this CSPN.

NGC~6778:
Previously classified as cont. by \citep{1994PASP..106...56F}.
\cite{2011A&A...531A.158M} showed
a high quality spectra where the absorptions at H$\beta$, \ion{He}{ii},
and \ion{N}{v}  (4603-19\AA) are
evident.
We classify this CSPN as O(H)3-4.

Pa~5:
\cite{2014AJ....148...57G} classified the central star of 
this new PN as PG~1159. 
They reported \ion{He}{ii}, Balmer series, and \ion{Ca}{ii} photospheric lines.
On the other hand \cite{2015MNRAS.448.3587D} classified the star as O(He).
We consider that the lines identified by these authors are insufficient 
to classify this object as a PG~1159. We prefer a spectral classification O(H).

PRTM~1:
\cite{1998ApJ...504L.103P}
reported that the central star have an
O(H) spectral type. Later, the high quality
spectra obtained by \cite{2012A&A...545A.146B}
display absorption lines of
\ion{He}{ii} (5412 and 4686\AA),
\ion{O}{v} (5114\AA), \ion{N}{v} (4603-19\AA),
and emission lines of
\ion{C}{iv} (4658, 5801-12\AA)
\ion{O}{v}  (4930\AA),
\ion{N}{v} (4945\AA).
All of this features are compatible with an O(H)3-4 type.
In this sense, it is surprising that the line \ion{N}{iv} at 5200-03\AA\ does not appear.
In addition, the authors identified an emission line at
5292\AA, that they associated to \ion{O}{vi}.
Nevertheless, the NIST Atomic Spectra Database
\citep{NIST_ASD}
do not include this line for \ion{O}{vi}.
Instead, we consider that this line better match with \ion{Ne}{iv}.

SkAc~1:
The CSPN of this object was identified in a WD survey
\citep{2016MNRAS.455.3413K}. 
Although the authors did not assign a spectral classification 
to the spectrum, the Balmer series
together with \ion{He}{ii} absorption at 4686 and 5412\AA are clearly seen.
We opt for a spectral classification DAO.

TC~1:
SECGPN reported a spectral type Of(H),
and in the spectra showed by \cite{1988A&A...190..113M} the \ion{He}{i} lines
are clear and strong. 
\cite{1986ApJ...301..772Z} reported P-Cygni profile. 
For this reason, we prefer a classification
O(H)5-9f.

TS~01:
\cite{2004ApJ...616..485T}
classified this CSPN as WD/NS, nevertheless in
the spectrum that these author showed, the Balmer series is 
clear but the \ion{He}{ii} stellar lines are not. 
With this evidence, we prefer a classification
H-rich.
\cite{2009PASP..121..316D} reported that the secondary star is a 
WD/NS or sdB.
Finally, \cite{2010ApJ...714..178T} showed a spectra with
H and \ion{He}{ii}, 4686\AA\ line, detectable in both emission and absorption.
According to the log g of the optical component, determined by 
\cite{2010ApJ...714..178T}, the CSPN is indeed a sdO star.

\end{appendix}

% las tablas del catalogo
%%%%%%%%%%%%%%%%%%%%%%%%%%%%%%%%%%%%%%%%%%%%%%%%%%%%%%%%%%%%%%%%%
%%%%%%%%%%%%%%%%%%%%%%%%%%%%%%%%%%%%%%%%%%%%%%%%%%%%%%%%%%%%%%%%%
%%%%%%%%%%%%%%%%%%%%%%%%%%%%%%%%%%%%%%%%%%%%%%%%%%%%%%%%%%%%%%%%%
%%%%%%%%%%%%%%%%%%%%%%%%%%%%%%%%%%%%%%%%%%%%%%%%%%%%%%%%%%%%%%%%%

\onecolumn
{\small
\begin{landscape}
\begin{longtable}{l c l c c c c c c c c}
\caption[Catalogue of CSPNe]{Catalogue of CSPNe. Description of the columns is in Section \ref{des-1}.}
\label{tabla1} \\

\hline  \multicolumn{1}{c}{PN G} & \multicolumn{1}{c}{f\_} & \multicolumn{1}{c}{Name} & \multicolumn{1}{c}{RA$_{J2000}$} & 
\multicolumn{1}{c}{DEC$_{J2000}$} & \multicolumn{1}{c}{SpT} & \multicolumn{1}{c}{Ref.} & \multicolumn{1}{c}{SpT2} & \multicolumn{1}{c}{Ref.} & 
\multicolumn{1}{c}{RB} & 
\multicolumn{1}{c}{Ref. spectra} \\ \hline 
\endfirsthead

\multicolumn{10}{c}%
{{\bfseries \tablename\ \thetable{} -- continued from previous page}} \\
\hline  \multicolumn{1}{c}{PN G} & \multicolumn{1}{c}{f\_} & \multicolumn{1}{c}{Name} & \multicolumn{1}{c}{RA$_{J2000}$} & 
\multicolumn{1}{c}{DEC$_{J2000}$} & \multicolumn{1}{c}{SpT} & \multicolumn{1}{c}{Ref.} & \multicolumn{1}{c}{SpT2} & \multicolumn{1}{c}{Ref.} & 
\multicolumn{1}{c}{RB} & \multicolumn{1}{c}{Ref. spectra}\\ \hline 
\endhead

\hline\multicolumn{10}{r}{{Continued on next page}} \\ 
\endfoot

\hline \hline
\endlastfoot
%
%1     &            & *   & KPD 0005+5106    & 00 08 18.4 & 51 23 19.0  & DO                & D1999                &             &              &    &                 \\
%5     &            & *   & PG 1034+001      & 10 37 04.0 & -00 08 20.0 & DO                & HI2004               &             &                  &  &                    \\
\#1  & * & KPD~0005$+$5106 & 00 08 18.4 & $+$51 23 19.0 & DO & D1999 &  &  &  & DL1985 DW1996 \\
\#2  & * & PG~0108$+$101 & 01 11 06.6 & $+$10 21 39.2 & DO & D1999 &  &  &  & WG1985 \\
\#3  & * & PG~0109$+$111 & 01 12 22.9 & $+$11 23 36.6 & DO & D1999 &  &  &  & WG1985 DW1996 \\
\#4  & * & PG~1034$+$001 & 10 37 04.0 & $-$00 08 20.0 & DO + ? & HI2004 &  &  & AL2019a & WG1985 WD1995 \\
\#5  & * & Hen~2$-$139 & 15 54 44.5 & $-$55 29 34.1 & B[e] & LZ1998 & SySt? & MZ2002 &  & MZ2002 \\
\#6  & * & 2MASS~18482\footnote{2MASS~J18482874$-$1237434} & 18 48 28.7 & $-$12 37 43.5 & O & t.w. &  &  &  & MM2014 \\
\#7  & * & Kn~61 & 19 21 39.0 & $+$38 18 56.9 & PG~1159 (lg E) & GD2014 &  &  &  & GD2014 \\
\#8  & * & PM~1$-$322 & 20 14 50.9 & $+$12 03 50.0 & SySt & MV2010 &  &  &  & - \\
\end{longtable}
\end{landscape}
}

%%%%%%%%%%%%%%%%%%%%%%%%%%%%%%%%%%%%%%%%%%%%%%%%%%%%%%%%%%%%%%%%%
%%%%%%%%%%%%%%%%%%%%%%%%%%%%%%%%%%%%%%%%%%%%%%%%%%%%%%%%%%%%%%%%%
%%%%%%%%%%%%%%%%%%%%%%%%%%%%%%%%%%%%%%%%%%%%%%%%%%%%%%%%%%%%%%%%%
%%%%%%%%%%%%%%%%%%%%%%%%%%%%%%%%%%%%%%%%%%%%%%%%%%%%%%%%%%%%%%%%%

{\tiny
\begin{landscape}
\begin{longtable}{cccccccccc}
\caption[Catalogue of CSPNe]{Physical parameters of CSPNe. Description of the columns is in Section \ref{des-2}.}
\label{tabla2} \\

\hline \multicolumn{1}{c}{PN~G} & \multicolumn{1}{c}{log $g$} & \multicolumn{1}{c}{ref.} & \multicolumn{1}{c}{met.} & \multicolumn{1}{c}{log $T$} & 
\multicolumn{1}{c}{ref.} & \multicolumn{1}{c}{log $(L_\star/L_{\odot})$} & \multicolumn{1}{c}{ref.} & \multicolumn{1}{c}{mag.} & \multicolumn{1}{c}{ref.}  \\ \hline 
\endfirsthead

\multicolumn{10}{c}%
{{\bfseries \tablename\ \thetable{} -- continued from previous page}} \\
\hline \multicolumn{1}{c}{PN~G} & \multicolumn{1}{c}{log $g$} & \multicolumn{1}{c}{ref.} & \multicolumn{1}{c}{met.} & \multicolumn{1}{c}{log $T$} & 
\multicolumn{1}{c}{ref.} & \multicolumn{1}{c}{log $(L_\star/L_{\odot})$} & \multicolumn{1}{c}{ref.} & \multicolumn{1}{c}{mag.} & \multicolumn{1}{c}{ref.}  \\ \hline 
\endhead

\hline\multicolumn{10}{r}{{Continued on next page}} \\ 
\endfoot

\hline \hline
\endlastfoot
\# 1 & 7.00 & N1999 & eff & 5.079 & N1999 & 3.700 & WW2010 & v  12.99 & WW2010 \\
\# 2 & 7.50 & N1999 & eff & 4.978 & N1999 & 1.000 & R1987  & v  15.60 & R1987 \\
\# 3 & 8.00 & N1999 & eff & 5.041 & N1999 & -     & -      & v  15.40$\pm$0.05 & ZF2012 \\
\# 4 & 7.50 & N1999 & eff & 5.000 & N1999 & - & - & v  13.21 & HD2007 \\
\# 5 & - & - &  & - & - & - & - & - & - \\
\# 6 & - & - &  & - & - & - & - & - & - \\
\# 7 & 8.30 & dO2015 & eff & 5.079 & dO2015 & - & - & r  18.48$\pm$0.01 & GD2014 \\
\# 8 & - & - &  & - & - & - & - & - & - \\
\end{longtable}
\end{landscape}
}

%%%%%%%%%%%%%%%%%%%%%%%%%%%%%%%%%%%%%%%%%%%%%%%%%%%%%%%%%%%%%%%%%
%%%%%%%%%%%%%%%%%%%%%%%%%%%%%%%%%%%%%%%%%%%%%%%%%%%%%%%%%%%%%%%%%
%%%%%%%%%%%%%%%%%%%%%%%%%%%%%%%%%%%%%%%%%%%%%%%%%%%%%%%%%%%%%%%%%
%%%%%%%%%%%%%%%%%%%%%%%%%%%%%%%%%%%%%%%%%%%%%%%%%%%%%%%%%%%%%%%%%

\end{document}